\documentclass[journal=jctcce,manuscript=article]{achemso}
\setkeys{acs}{articletitle = true}

\usepackage{todonotes}
\usepackage{graphicx}%
\usepackage{dcolumn}%
\usepackage{bm}%
\usepackage[normalem]{ulem}
\usepackage{microtype}
\usepackage{xcolor}
\usepackage{amsmath,amssymb,amsfonts,mathtools,cancel}
\usepackage{multirow}
\usepackage{subcaption}

\definecolor{DragonGreen}{RGB}{0,126,48}
\definecolor{DarkOrchid}{RGB}{153,50,204}
\newcommand{\mc}[1]{{ #1 }}

\definecolor{Brownie}{RGB}{97,16,9}

\newcommand{\bx}{\ensuremath{\mathbf{x}}}
\newcommand{\bmu}{\ensuremath{\boldsymbol{\mu}}}
\newcommand{\by}{\ensuremath{\mathbf{y}}}

\newcommand{\Tr}{\ensuremath{\operatorname{Tr}}}

\title{
Recognizing Local and Global \\ Structural Motifs At the Atomic Scale
}%

\author{Piero Gasparotto}
 \affiliation{Laboratory of Computational Science and Modeling, Institute of Materials, {\'E}cole Polytechnique F{\'e}d{\'e}rale de Lausanne, 1015 Lausanne, Switzerland}%
 
\author{Robert Horst Mei{\ss}ner}
 \affiliation{Laboratory of Computational Science and Modeling, Institute of Materials, {\'E}cole Polytechnique F{\'e}d{\'e}rale de Lausanne, 1015 Lausanne, Switzerland}%
 
\author{Michele Ceriotti}
\email{michele.ceriotti@epfl.ch}
\affiliation{Laboratory of Computational Science and Modeling, Institute of Materials, {\'E}cole Polytechnique F{\'e}d{\'e}rale de Lausanne, 1015 Lausanne, Switzerland}%

\date{\today}%
\begin{document}
\begin{abstract}

Most of the current understanding of structure-property relations at the molecular and the supramolecular scales 
can be formulated in terms of the stability of and the interactions between a limited number of recurring structural motifs (e.g. H-bonds, coordination polyhedra, protein secondary-structure). 
Here we demonstrate an algorithm to automatically recognize such patterns, based on the identification of local maxima in the probability distributions observed in atomistic computer simulations, which is robust to the dimensionality and the sparsity of the reference atomistic data. 
We first discuss its main features, demonstrating some on artificial datasets, and then show how it can be applied to identify coordination environments in Lennard-Jones clusters, and to recognize secondary-structure patterns in the simulation of an oligopeptide. 
To assess the applicability of this algorithm for motifs that involve several interdependent degrees of freedom, we also employ it to identify groups of conformers of the cluster and the polypeptide, considered in their entirety. 
The motifs identified by analyzing atomistic simulations can be used to interpret and rationalize the stability and behavior of the system at hand, and also as a tool to accelerate sampling, in association with biased molecular dynamics schemes. 

\end{abstract}
\maketitle

\section{Introduction}

The amount, availability and performance of computational facilities dedicated to atomic-scale modelling have increased dramatically in the last decades \cite{fisc2006natmat,marzari2016materials}.
As a consequence, trajectories of several microseconds \cite{lind+16jpcb}, and systems comprising million of atoms \cite{zhao+13nat} have become possible in atomistic simulations, and computationally-generated databases of materials properties have emerged as a tool to accelerate the design and discovery of new materials with improved properties~\cite{agra16apl,bere16aplm,curt+13nat,phil16aplm,pizzi2016}.
The tremendous increase of the amount and complexity of data produced from simulations poses a challenge of its own, particularly when it comes to recognizing the molecular patterns that underlie the behavior of large, slowly-evolving systems. 
The traditional paradigm in which simulations were manually inspected, and analyzed in terms of structural descriptors designed by trial and error is becoming impractical:
the amount and complexity of potential correlations concealed in such an amount of data are far from being fully understandable by a human. %
A promising approach to facilitate the interpretation of large, complex databases and trajectories involves adapting established data-mining techniques to the analysis of atomic-scale data.
For instance, one could apply clustering approaches to identify (meta)stable configurations, and partition the free-energy landscape in a set of fluctuation basins associated with each configuration. These basins can then be used as the basis for coarse-grained descriptions of the dynamics, such as Markov state models~\cite{chod07jcp,de01jmb,andr05pnas,sing04jcp}.
Many successful examples of machine-learning (ML)-based methods applied to the study of materials have also appeared in the literature in recent years \cite{huan+15prb,rupp12prl,fabe16prl,behl16jcp,de17jchmi,spar16scrm}.
A particularly intriguing approach attempts to identify not only stable configurations of the system, but also the elementary structural motifs that, alone or combined in non-trivial ways, determine the stability and properties of complex supramolecular structures. This idea
has seen application in the context of protein structure\cite{rost+94prot,ball2010mach,chen06bion,dale10bioi} materials\cite{hara+10jctc,carr09mmm,schu14prb} and molecular systems \cite{kaya12jcim}.
Recently the Probabilistic Analysis of Molecular Motifs (PAMM) algorithm has been developed with the precise goal of identifying molecular patterns based on an analysis of the probability distribution of fragments observed in an atomistic simulation~\cite{gasp-ceri14jcp}.
PAMM has this far been used to introduce a data-driven, agnostic definition of fundamental, yet simple, entities such as the hydrogen bond~\cite{gasp-ceri14jcp}, to recognize defects and correlations in liquid water~\cite{gasp+16jctc}, and to single out protonated water species based on an analysis of their electronic structure~\cite{ross16jpcl}.

In this paper we address some of the limitations of the PAMM scheme, namely the stability with respect to dimensionality and sparse data sampling, and the reliability in case of periodic inputs or non-Gaussian features. In particular, we introduce an analysis of the stability of PAMM clusters that makes it possible to recognize the hierarchical structure of the free energy landscape. We show how the improved scheme can be used successfully not only to recognize local environments and molecular patterns, but also be extended to identify global structural features.
In Section I we discuss the rationale and the practical implementation of our methodology.
In Section II we apply our method to two realistic datasets. First, we identify coordination environments in a Lennard-Jones cluster and then recognize secondary-structure patterns in a $\beta$-hairpin oligopeptide.
In Section III, as a demonstration of the stability of our approach when applied to high-dimensional descriptors, we use PAMM to perform a classification of entire structures, and show how clustering techniques can be combined with non-linear dimensionality reduction schemes to fully-characterize the configuration space of complex atomistic systems.
Finally we draw our conclusions and discuss possible applications of probabilistic motif identifiers in the context of accelerated sampling  schemes.

\section{Probabilistic analysis of Molecular Motifs}

\begin{figure*}[bth]
\centering
\includegraphics[width=1.0\textwidth]{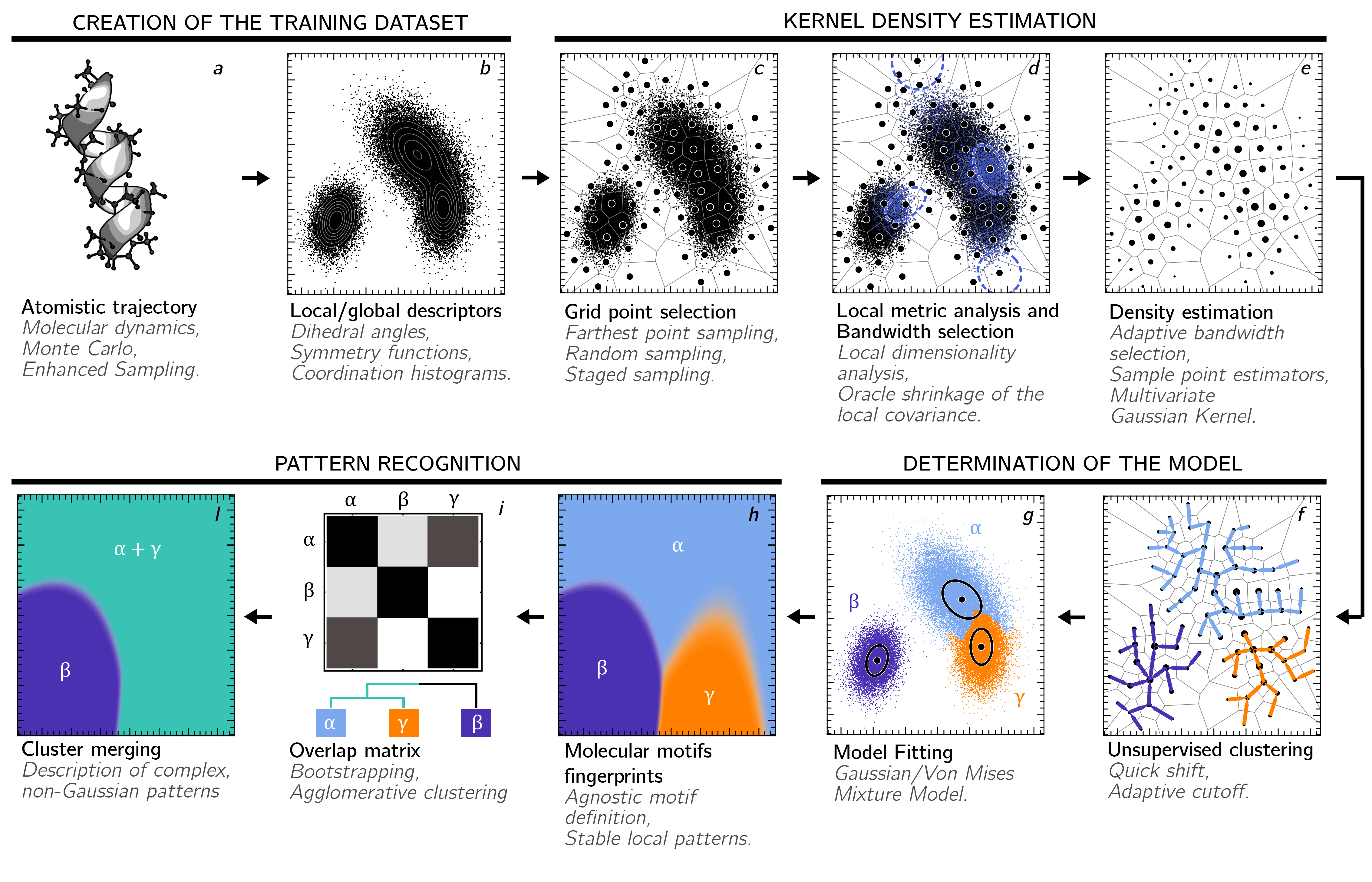}
\caption{Schematic illustration of the PAMM workflow applied to an artificial two-dimensional dataset.}
\label{fig:pamm_flow}
\end{figure*}

The clustering algorithm we discuss in this paper follows the same philosophy of the probabilistic analysis of molecular motifs (PAMM) method~\cite{gasp-ceri14jcp}, including however a number of crucial additional ingredients to address the limitations of the original scheme. 
PAMM aims to provide an unambiguous and unbiased approach to recognize molecular motifs apparent in data produced from atomistic simulations, by employing machine-learning techniques.
Since the distribution of the structural descriptors reflects the (free) energetic stability of the molecular patterns, we use the probability distributions observed in the simulations to inform the identification of clusters.
In line with the free-energetic interpretation, we choose to characterize distinct clusters as separate modes of the distribution, corresponding to the basins of attraction of local maxima. 

Our approach can be split into several steps, that are summarized in Fig.~\ref{fig:pamm_flow}.
We will discuss separately the details of the algorithms we use for each step, which we found to be capable of providing robust clustering in a variety of realistic atomistic simulation problems.
Different choices of course would be possible, as we will discuss when appropriate.

\subsection{Input Representation}

For most applications, it is impractical to directly use the Cartesian coordinates of the atoms to describe structures or local environments. One should rather represent them in terms of a (possibly large) number of ``order parameters'' or ``fingerprints'', that provide an unbiased and sufficiently complete description of the geometry while fulfilling all of the important symmetries -- such as being invariant to atom labelling or to rigid rotations and translations~\cite{behl-parr07prl,bart+13prb,sade+13jcp}. 
Seen through this lens, the atomistic simulation data is converted to a set $\mathcal{X}=\left\{\mathbf{x}_i\right\}$ containing a large number $N$ of $D$-dimensional vectors, $\mathbf{x}_i\in \mathbb{R}^D$, with each vector representing either a subset or the entirety of the atoms within a structure.
\mc{It is worth stressing that the selection of a proper set of descriptors is far from being a trivial point, and can influence deeply the outcome of the subsequent analysis. To mitigate this problem, we optimized the different ingredients in PAMM to be robust with growing dimensionality, so that many order parameters can be used simultaneously to deal e.g. with heterogeneous systems. 
One could also use more abstract descriptors that are based on more or less systematic expansions of atomic environments -- for instance Behler-Parrinello symmetry functions~\cite{behl11jcp}, SOAP power spectra~\cite{bart+13prb}, SPRINT coordinates~\cite{piet-andr11prl}. 
The choice of input order parameters is also important, because it determines the metric relative to which probability distributions and free energies are computed~\cite{KITAO1999164,branda2007}. A poor choice can generate spurious maxima in the probability distribution, or merge kinetically separate states into a single basin. These artifacts can be corrected, at least in part, by taking into account kinetic information in the definition of the order parameters~\cite{nuske2014variational,tiwary2016spectral}.
}

\subsection{Grid Selection}
In order to mitigate the high cost of density estimation for large datasets in high dimension, the first step in the PAMM workflow involves the selection of a sparse grid $\mathcal{Y}\subseteq \mathcal{X}$ 
containing $M\ll N$ points.
A grid covering almost uniformly the parameter space spanned by $\mathcal{X}$ can be obtained using a greedy farthest-point sampling procedure~\cite{rosenkrantz+77siam}, that can be modified to use a finer grid in the high-probability regions~\cite{ceri+13jctc}. The computational cost of this selection is $\mathcal{O}(MN)$, so it can be performed also on very large data sets. The determination of the grid by sub-sampling the full set $\mathcal{X}$ also allows one to partition data into neighborhoods of the grid. For instance, one can construct the Voronoi polyhedra for $\mathcal{Y}$, and assign each datum $x$ to the Voronoi set $\mathcal{V}_i$ of the closest-by grid point $y_i$.
Different strategies for subsampling are also possible~\cite{ceri+10jctc,Prabhakaran2012}. It should be stressed, however, that the subsequent steps in the PAMM workflow are designed to minimize the impact of the grid size on the final outcome, and to guarantee that in the limit $M,N\rightarrow \infty$ there is no dependence at all. 

\subsection{Kernel Density Estimation}

Density-based clustering algorithms~\cite{ester1996density,rodr-laio14science} depend crucially on the quality of the estimation of the underlying probability density.
Kernel-density estimation (KDE) provides a smooth, robust approach to do so, that also leaves the flexibility to adapt to strongly anisotropic probability distributions and/or non-Euclidean geometries.

The KDE on a grid point $\mathbf{y}_i$ can be written as 
\begin{equation}
  P(\mathbf{y}_i) = \frac{1}{\sum_{j=1}^N w_j} \sum_{j=1}^N w_j K_{\mathbf{H}_j}(\mathbf{x}_j - \mathbf{y}_i),
  \label{eqn:kde}
\end{equation}
where we use a very general expression in which each data point can be assigned a weight $w_i$ (e.g. to compensate for biased sampling), and an adaptive bandwidth matrix $\mathbf{H}_j$. The use of a grid implies that (for a fixed grid) the cost of evaluating $P$ scales only linearly with the number of data points.

We use an anisotropic multivariate Gaussian kernel,
\begin{equation}
  K_{\mathbf{H}}(\mathbf{x}) = \frac{1}{\sqrt{(2\pi)^D |\mathbf{H}|}} \exp\left[ -\frac{1}{2}\mathbf{x}^T\mathbf{H}^{-1}\mathbf{x} \right],
  \label{eqn:kernel}
\end{equation}
that provides enough flexibility to adapt to strong variations of the geometric distribution of data points.
A common problem with kernel density estimation, which is particularly severe for high-dimensional and/or sparsely populated datasets, is the optimization of the bandwidth of the kernel. 
The shape of the kernel, encoded in the bandwidth matrix $\mathbf{H}$ in eq. \eqref{eqn:kernel}, determines a trade-off between the statistical noise in the estimated density and a systematic error due to the smoothing of the true underlying density.
When the true density $P^\star(\mathbf{x})$ is known, the optimal bandwidth can be selected by minimizing the Mean Integrated Squared Error (MISE, $\int \mathrm{d}\mathbf{x}\left[P^\star(\mathbf{x})-P(\mathbf{x})\right]^2$). 
In general, obviously, the true density is not known, and one has to resort to recipes to choose the bandwidth that are derived based on some reasonable assumptions about the underlying distribution.
A particularly simple heuristic for selecting the bandwidth is given by Silverman's rule~\cite{scott+92kde,bow+97app},
\begin{equation}
  \mathbf{H} = \left[ \frac{4}{N(D+2)} \right]^{\frac{2}{D+4}} {\boldsymbol{\Sigma}},
  \label{eqn:scotts_rule}
\end{equation}
where ${\boldsymbol{\Sigma}}$ is the covariance of the entire dataset, $N$ the number of data points and $D$ the dimensionality. 

\subsubsection{Local Kernel Metric Analysis}

Silverman's rule~\eqref{eqn:scotts_rule} minimizes the MISE for a single multivariate normal distribution. For a general distribution, composed of many separate peaks, this generally results in an over-estimation of the bandwidth, and in loss of resolution unless an extremely large amount of data is available. 
As a possible solution, and to provide a mechanism to fine-tune the balance between resolution and statistical noise, we propose a simple strategy to localize the determination of the bandwidth and of the dimensionality of the data. 
Basically, the idea is to apply Silverman's rule to subsets of the full dataset.

In order to estimate the optimal KDE bandwidth for a sample $\bx_i$, we introduce weighting factors for each of the other sample points $\bx_j$ around it, that are computed from a spherical Gaussian
\begin{equation}
u_{ij} = \exp\left[-\frac{(\bx_j-\bx_i)^2}{2\cdot \sigma_i^2}\right] N w_j/\sum_j w_j, \label{eq:loc-weights}
\end{equation}
where $\sigma_i$ is a localization factor whose choice we will discuss below. The weights-adjusted sample population for the selected point is computed as $N_i=\sum_j u_{ij}$.
We find it convenient to introduce two possible approaches to determine the localization parameters $\sigma_i$. (i) in cases where one expects the spatial extent of clusters to be relatively homogeneous, one can choose a fixed localization window expressed as a fraction of the overall spatial extent of the dataset, assessed as the global covariance matrix of the data, $\mathbf{\Sigma}$; this can be achieved setting  $\sigma_i=f_\text{s} \sqrt{\Tr\mathbf{\Sigma}}$. In this case, each localization region can contain a different weight-adjusted population $N_i$. 
(ii) In cases where one expects clusters with very different spreads, but similar populations, it might be more convenient to use a position-dependent localization window that is adjusted so that each region contains a prescribed fraction $f_\text{p}$ of the total number of weighted data points. Each $\sigma_i$ should then be adjusted iteratively until $N_i\approx N f_\text{p}$. A quantitative discussion of the relative merits of the two strategies is reported in the SI (Fig. S1).

The local weights from eq.~\eqref{eq:loc-weights} can be used, for instance, to estimate the local covariance $\mathbf{\Sigma}_i$, around each data point.
Each element of the covariance is estimated by
\begin{equation}
    \left[\mathbf{\Sigma}_i\right]_{kl} = \sum_{j=1}^{N}{u_{ij}\cdot(x_{jk}-\bar{x}_k)(x_{jl}-\bar{x}_l)} / N_i \label{eq:local-sigma}
\end{equation}
where $\bar{\mathbf{x}}=\sum_{j=1}^N{u_{ij} \mathbf{x}_i}/N_i$. 
Computing the covariance of a subset of the points can exacerbate stability problems that are also present, in general, when the sampling is insufficient or when different degrees of freedom are strongly correlated. In these circumstances, $\mathbf{\Sigma}_i$ can be very ill-conditioned, and even have eigenvalues that are zero to within machine precision. %
This is a consequence of the fact that the usual estimator for the covariance $\mathbf{\Sigma}_i$ is a biased estimator of its inverse $\mathbf{\Sigma}_i^{-1}$. This is a well-known problem that is typically addressed by introducing alternative estimators that are less strongly biased. 
Here we use the so-called Oracle Approximating Shrinkage (OAS) estimator \cite{chen+10ieee} that reads
\begin{equation}
    \widetilde{\mathbf{\Sigma}}_i = (1-\psi_i)\mathbf{\Sigma}_i + \frac{\psi_i\mathrm{Tr}(\mathbf{\Sigma}_i)\mathbf{I}}{D}
\end{equation}
where 
\begin{equation*}
  \psi_i = \min\left[1, \frac{\left(1-\frac{2}{D}\right)\mathrm{Tr}(\mathbf{\Sigma}_i^2)+\mathrm{Tr}^2(\mathbf{\Sigma}_i)}{\left(N_i+1-\frac{2}{D}\right)\mathrm{Tr}(\mathbf{\Sigma}_i^2)-\frac{\mathrm{Tr}^2(\mathbf{\Sigma}_i)}{D}}\right].
\end{equation*}

Furthermore, the eigenvalue spectrum of the local covariance matrix can be used to estimate an effective local dimensionality $D_i$ based on the effective rank of $\mathbf{\Sigma}_i$ \cite{roy+07espc}:
\begin{equation}
  D_i = \exp\left( -\sum_{k=1}^D \eta_k \log(\eta_k) \right)
\end{equation}
where $\eta_k = \lambda_k / \sum_{k=1}^D |\lambda_k|$ and $\left\{\lambda_{k}\right\}$ is the eigenvalue spectrum of $\mathbf{\Sigma}_i$.

Given the local covariance matrices and an estimate of the local dimensionality, one can introduce an expression for the optimal bandwidth matrices to be used in the KDE.
Assuming that each local zone resembles a normal distribution, the optimal bandwidth for $\bx_i$ can finally be obtained as a localized version of Silverman's Rule~(\ref{eqn:scotts_rule}):
\begin{equation}
  \mathbf{H}_i = \left[ \frac{4}{N_i(D_i+2)} \right]^{\frac{2}{D_i+4}} \widetilde{\mathbf{\Sigma}}_i.
\end{equation}
Performing this analysis for each datum would entail poor scaling of the procedure with the total number of points. One can however exploit the definition of a sparse grid to accelerate greatly the computation, without changing the spirit of the localization strategy. 
First, one can compute the local bandwidth $\mathbf{H}_i$ only for the grid points $\by_i$, and assign it to all the samples that belong to its Voronoi set,  i.e. $\mathbf{H}_j\equiv \mathbf{H}_i$, $\widetilde{\mathbf{\Sigma}}_j \equiv \widetilde{\mathbf{\Sigma}}_i \ \ \forall \bx_j \in \mathcal{V}_i$.
Furthermore, the evaluation of eq.~\eqref{eq:local-sigma} can be accelerated by including the contribution from grid points beyond a reasonable cutoff distance using only the position of the grid $\by_i$ and assigning to it a weight proportional to the total weight of points within its associated Voronoi polyhedron, rather than summing over all sample points associated to it. 

 Finally, we note that in cases in which sampling is particularly irregular, it can happen that the bandwidth is smaller than the distance with the nearest neighbor of a grid point. Often these outliers result from insufficient grid size and/or non-Gaussian tails of the distribution, and should not generate additional clusters. To avoid this, we increase automatically the bandwidth to match the first-neighbor distance, but issue a warning to allow for manual inspection to determine whether the outlier is of some significance. 

\subsection{Identification of Motifs}

After having estimated the density at the grid points $P(\mathbf{y}_i)$, one can proceed to use it to subdivide the distribution into several distinct clusters. As discussed above, we chose to identify clusters that represent recurring molecular patterns as maxima in the probability distribution, and to associate to each maximum all the grid points falling within its basin of attraction.
In atomistic trajectories this construction has a profound physical interpretation: each identified maximum of the probability distribution can be associated with a free energy (meta-)stable minimum of the $D$-dimensional description of a group of atoms.

We perform a non-parametric clustering based on this idea using the Quick-Shift algorithm \cite{vedaldi+08cv}. \mc{Other clustering algorithms could obviously be used - see the SI for a comparison of some of the most common methods~\cite{carre00mode,macq67kmeans,este96dbscan}.} 
Starting from a random grid point which has not been assigned yet to a  cluster, one connects it to the nearest grid point that has a higher probability density, i.e., $\mathbf{y}_i$ is connected to $\mathbf{y}_j$ such that
\begin{equation}
    j = \underset{P(\mathbf{y}_i)>P(\mathbf{y}_j)}{\mathrm{argmin}}|\mathbf{y}_i-\mathbf{y}_j| .
\end{equation}
The procedure can be interrupted based on a suitable stopping criterion, as discussed below. The final point is identified as a maximum, and tagged as the center $\mathbf{z}_k$ of a cluster.  All the points in the chain are tagged as belonging to the associated set $\mathcal{Z}_k$. 
One can then start climbing from another unassigned point, and stop when the procedure encounters another maximum, or one of the points that have already been assigned to one of the $\mathcal{Z}_k$ clusters, with which the current chain would then be merged.
After all points have been traversed, the grid $\mathcal{Y}$ will be partitioned into $n$ disjoint sets, without the need of specifying \emph{a priori} the number of clusters or their geometry. 
The stopping criterion for Quick-Shift is typically set by requiring that the Euclidean distance between the current point and the next one in the chain is below a set threshold $\Delta$.
In line with the spirit of the local metric analysis we perform to obtain the bandwidth matrix for the KDE, we select the cutoff locally. At each stage in the procedure, the cutoff is chosen based on the covariance associated with the grid point that is being considered, i.e.
\begin{equation}
    \Delta_i = \alpha \sqrt{ \operatorname{Tr} \widetilde{\mathbf{\Sigma}}_i }.
    \label{eqn:qs_scale}
\end{equation}
This choice adapts the cutoff to the local spread of the data, and is consistent with the Gaussian assumption that underlies the localization procedure. For a multivariate Gaussian, it would cluster together two points that are drawn at random from the distribution.
If needed, the values of $\Delta_i$ can be further adjusted by a multiplicative factor $\alpha$, to fine-tune the resolution of the clustering procedure. In the examples we used to benchmark this method we found only small changes in the final clustering upon scaling the cutoff by $\pm10$\,\%.

\subsubsection{Gaussian Mixture Models}

After having determined the number and position of modes of the distribution, we fit a Gaussian Mixture Model (GMM) to the data. This step is useful as it provides a simple interpretation of the density in terms of a number of separate modes, making it possible to develop fingerprints for the different molecular motifs that have a transparent probabilistic interpretation.
The probability distribution is thus fitted to a sum of $n$ multivariate Gaussians
\begin{equation}
  \hat{P}(\mathbf{x}) = \sum_{k=1}^{n} p_k G(\mathbf{x}|\bmu_k,\mathbf{\Sigma}_k),
\end{equation}
Each Gaussian, associated with a weight $p_k$, is defined as
\begin{equation}
  G(\mathbf{x}|\bmu,\mathbf{\Sigma}) = \frac{1}{\sqrt{(2\pi)^D \mathrm{det}\mathbf{\Sigma}}} e^{-(\bx-\bmu)^T\mathbf{\Sigma}^{-1}(\bx-\bmu)/2}\label{eq:gmm-gauss}
\end{equation}
where $\mathbf{\Sigma}$ is its covariance matrix and $\boldsymbol{\mu}$ its mean position.
Rather than fitting the Gaussian parameters with an expectation-maximization algorithm, we exploit the fact that we know the number $n$ and modes $\mathbf{z}_k$ of clusters. We set the mean of the Gaussian cluster to the mode of the cluster, and estimate the covariance with the usual expression:
\begin{equation}
  \begin{split} 
    \bmu_k = \mathbf{z}_k , \quad p_k = \sum_{\by \in \mathcal{Z}_k} P(\by) \big/ \sum_{\by \in \mathcal{Y}} P(\by), \quad  \\ 
    \bar{\by}_k = \sum_{\by \in \mathcal{Z}_k}  \by  P(\by)/ \sum_{\by \in \mathcal{Z}_k} P(\by) \\ 
    \mathbf{\Sigma}_k = \sum_{\by \in \mathcal{Z}_k}(\by-\bar{\by}_k) (\by-\bar{\by}_k)^T   P(\by)  \big/ \sum_{\by \in \mathcal{Z}_k} P(\by).
  \end{split}
  \label{eqn:gmm_parm}
\end{equation}

The reason for introducing this additional step, after having already obtained a non-parametric clustering by quick-shift, is that a GMM lends itself quite naturally to a probabilistic interpretation. Given a configuration associated with the fingerprints $\bx$, the expression
\begin{equation}
  \hat{P}_k(\bx) = p_k G(\mathbf{x}|\bmu_k,\mathbf{\Sigma}_k) / \left( \zeta+\hat{P}(\bx)\right)
  \label{eqn:gmm_partitioning}
\end{equation}
corresponds to the probability that such configuration belongs to the $k$th cluster. 
Eqn.~\eqref{eqn:gmm_partitioning} can therefore be used to introduce some smooth, fuzzy Probabilistic Motif Identifiers (PMI) that constitute a data-driven definition of a molecular pattern -- such as the hydrogen bond~\cite{gasp-ceri14jcp,gasp+16jctc} or the accumulation of charge around an excess proton~\cite{ross16jpcl}.
The ``background'' parameter $\zeta$ -- that defaults to zero and should in any case be set to a very small value -- serves to provide a more physical description of outlier configurations. In practice, configurations for which all Gaussian densities are below $\zeta$ are considered to be new, unclassified states that had not been sampled properly in the initial dataset.

\subsubsection{Periodic data}

When working with descriptors that are periodic in nature (e.g. angles and dihedrals) the probabilistic description should be adapted to account for the non-Euclidean geometry of the space. 
Von Mises distributions~\cite{mard+00dirstat,mard+08jcs} are the equivalent of a Gaussian on a circle, and can be used to describe periodic data, as they are smooth across the boundaries. 
The multivariate extension of a von Mises distribution, however, cannot be normalized analytically, which makes it hard to use it for the KDE step in our procedure. Furthermore, when the bandwidth is negligible with respect to the periodicity, a Gaussian kernel computed while using a minimal image convention in defining distances between points is virtually indistinguishable from a von Mises distribution. For this reason, we use multivariate Gaussian kernels in the KDE step, also along periodic directions. 

When determining the GMM that underlies our fingerprints, however, one cannot assume that the covariance associated with each cluster is small with respect to the periodicity of the pattern space. Given the difficulties with normalizing a multivariate von Mises distribution, we use a product of one-dimensional distributions to construct basis functions for the GMM that represent each cluster, i.e. we use 
\begin{equation}
  G(\bx|\bmu,\boldsymbol{\kappa}) = \prod_i^D \frac{e^{\kappa_i\cos(x_i-\mu_i)}}{2\pi I_0(\kappa)},
\end{equation}
in lieu of Eq.~\eqref{eq:gmm-gauss}, where $I_0$ is the modified Bessel function of order zero. The mean value and weight for each cluster are determined according to \eqref{eqn:gmm_parm}, whereas $\kappa_i$ are obtained using the conventional estimators for the concentration parameter of a one-dimensional von Mises distribution \cite{sra12cs}.

\begin{figure*}[bth]
\centering
\includegraphics[width=1.0\textwidth]{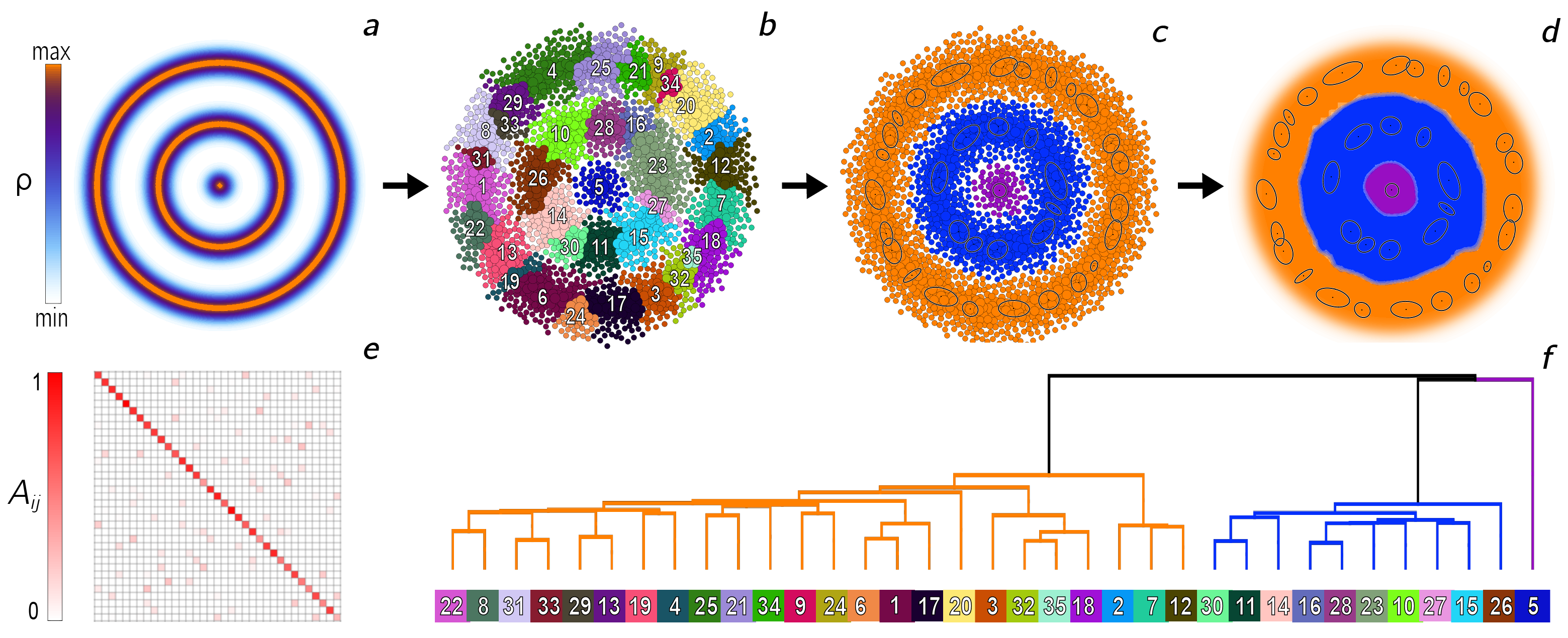}
    \caption{
    (a) Radially-symmetric probability density function that corresponds to concentrical circles. Clusters before (b) and after (c) merging and the PMIs corresponding to the final macroclusters (d). Panel (f) represents the dendrogram resulting from the agglomerative clustering based on the  cluster stability matrix $\mathbf{A}$ (e) and using a single-linkage strategy for the merging.
    }
    \label{fig:interlocked_circles}
\end{figure*}

\subsection{Error Assessment}

Computing the \emph{absolute} MISE requires knowledge of the underlying probability distribution.
It is however possible to estimate the statistical error associated with a given estimate, which can be useful to fine-tune the KDE parameters and to decide on the statistical significance of clusters that are identified in later stages of the PAMM analysis.
Bootstrapping provides a very general and well-established approach to infer the statistical error in a distribution, when its analytic form is not known \cite{efro79as}.
Bootstrapping relies on the analogy existing between the population and the sample drawn from it and consists in re-sampling with replacement a large number of sets from the given data, in order to build empirically an estimate of the probability distribution related to a certain statistical estimate.
In this specific case, we exploit bootstrapping to generate $N_\text{BS}$ independent samples of the KDE,  $P^{(m)}(\by_i)$. From these, one can estimate the standard error $\delta P(\by_i)$ associated with the KDE at each grid point.

The bootstrapping procedure is not only useful to get an estimate of the statistical error in the KDE. By performing the (deterministic) clustering procedure we discussed above on the $m$-th bootstrapped estimate of $P(\by_i)$, one can obtain clusters $\mathcal{Z}_k^{(m)}$ that reflect the statistical fluctuations of the KDE. 
The comparison between the bootstrapped clusters and those obtained on the straightforward KDE can then be used to compute indicators of how ``stable'' the clustering procedure can be considered. 
To do so, one can first compute 
\begin{equation}
\begin{split}
Q = \sum_i P(\by_i), \quad
Q_k = \sum_{y\in \mathcal{Z}_k} {P(\by)},\\
Q^{(m)}_k = \sum_{\by\in \mathcal{Z}^{(m)}_k} {P(\by)}{},\quad
Q^{(m)}_{j|k} = \sum_{\by\in \mathcal{Z}_k\cap\mathcal{Z}_j^{(m)}} \frac{P(\by)}{Q^{(m)}_k},
\end{split}
\end{equation}
and then introduce
\begin{equation}
A_{ij}= \frac{1}{N_\text{BS}\sqrt{Q_i Q_j}} \sum_m \sum_k Q^{(m)}_k Q^{(m)}_{i|k}Q^{(m)}_{j|k}.\label{eq:adjacency}
\end{equation}
For each bootstrapping run, this expression determines what is the probability that -- taking one of the bootstrap cluster at random, and drawing two sample points from them -- one would be part of the reference cluster $i$, and one of the reference cluster $j$, renormalized over the probabilities of clusters $i$ and $j$.
The diagonal elements $A_{ii}$ report on how robust is the determination of the $i$-th cluster. If the $i$-th cluster appears identical in each bootstrapping run, $A_{ii}$ takes a value of one, which becomes smaller if in one or more of the iterations the cluster is split in multiple clusters. Off-diagonal terms report on how ``fuzzy'' are the borders of the clusters. If no bootstrapping run generates a cluster that overlaps with both $\mathcal{Z}_i$ and $\mathcal{Z}_j$, $A_{ij}$ would take a value of zero, which increases if the clusters get merged in some of the runs, or if some of the $\mathcal{Z}_k^{(m)}$ include points from both clusters. 

\subsubsection{Cluster Association and Non-Gaussian Patterns.}
\label{chp:cluster_merge}

The cluster stability matrix $A_{ij}$ from eq.~\eqref{eq:adjacency}, can also be used to perform an additional ``meta-clustering'' step, that suggests ways to group together some of the clusters identified in the previous steps of PAMM, based on the notion that they were separated due to statistical error rather than because they correspond to separated free-energy basins. We attempted two approaches, that provide satisfactory and similar results, but differ in the underlying interpretation. 
One possibility is to choose a threshold value for the adjacency matrix, and find the connected components of the associated graph. This approach corresponds to a sort of ``flooding'' scheme, in which clusters that are above a prescribed level of fuzziness are merged at once.

Alternatively, one can proceed with a hierarchical clustering procedure~\cite{murt-cont12algorithms}, which can also be represented in a tree-like plot which helps interpreting the relations between different clusters~\cite{de+17jci}. Based on the adjacency matrix one can define a distance between clusters as $d_{ij}=-\log( A_{ij}/\sqrt{A_{ii} A_{jj}})$. The  pair of clusters which are closest is merged first, after which the merging is repeated iteratively until a single cluster remains. The cluster hierarchy can be represented as a binary tree, in which the vertical position of the branching point correspond to the distance between the leaves. Different strategies exist -- and can be tried to improve the resolving power of the method -- to define a distance between merged clusters. In this work we always use Ward's minimum-variance prescription~\cite{ward63jasa}, unless otherwise specified. This second strategy is more consistent with an interpretation in which fuzzy clusters correspond to clustering errors, and performs as little merging as possible to achieve the desired number of clusters, or degree of separation. 

\begin{figure*}[bth]
    \centering
    \includegraphics[width=\linewidth]{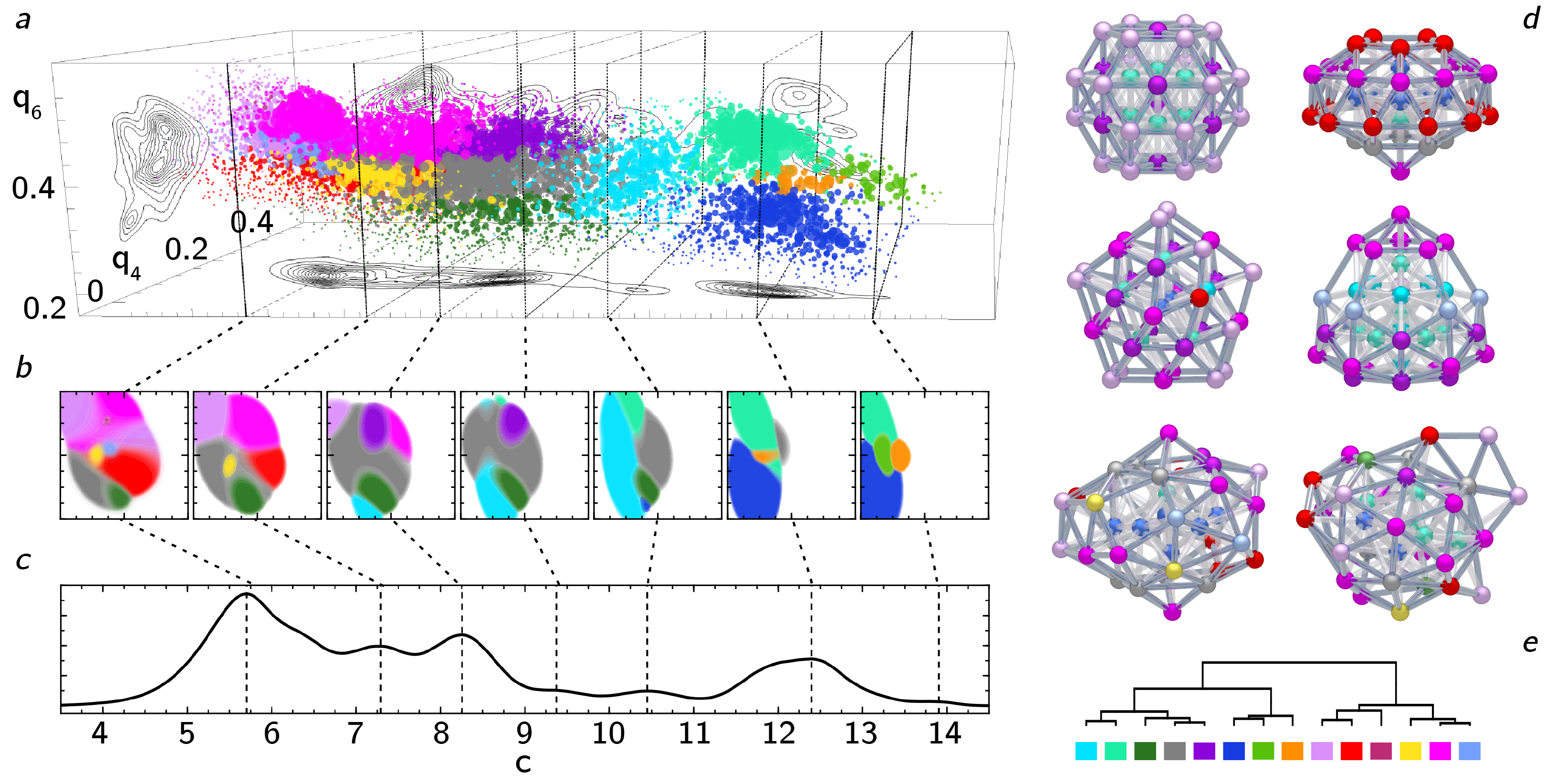}
\caption{A PAMM analysis of local environments in a LJ$_{38}$ cluster simulated at $T^\star=0.18$. (a) The probability distribution in fingerprint space, that include a smooth coordination number $c$, and the Steinhardt order parameters $q_4$ and $q_6$. Grid points have an area proportional to the KDE of the probability density, and are colored according to the quick-shift clustering. Marginal probabilities are also drawn as contour plots. (b) Slices of the probabilistic motif indicators that result from the GMM built on the PAMM clusters. Each cluster is indicated with the corresponding solid color when its PMI is equal to $1$, while the opacity linearly decreases to $0$ when the value of the PMI is $0$. (c) The probability distribution for $c$ only. (d) Representative configurations of the LJ$_{38}$ clusters, with atoms colored according to the dominant PMI. (e) Binary tree representation of the hierarchical clustering based on the adjacency matrix of the PAMM clusters.}
    \label{fig:lj38-env}
\end{figure*}

Besides making the clustering procedure more robust, this merging step is also useful to address the presence of strongly non-Gaussian features in fingerprint space. Even though the KDE and Quick-Shift algorithms are fully non-parametric, at many steps in our protocol we invoked the assumption that data can be (locally) described by multivariate Gaussian distributions.
As a specific, and rather extreme, example of non-Gaussian behavior, let us consider the distribution depicted in Fig.  \ref{fig:interlocked_circles}, corresponding to three concentric rings.
Figure~\ref{fig:interlocked_circles} demonstrates how, in the presence of non-Gaussian clusters, the partitioning of the data by PAMM is highly unstable, leading to an adjacency matrix showing considerable overlap between different clusters.
Hierarchical clustering, illustrated using a dendrogram in Fig. \ref{fig:interlocked_circles}e, shows clearly that there are three ``macro-cluster'' that correspond to the rings, while Gaussian features within each ring are clearly detected as being strongly connected. 
It is worth noting that, once different Gaussian clusters have been joined based on the adjacency matrix, it is easy to develop non-Gaussian fingerprints, by simply summing over all GMM fingerprints associated with each macro-cluster $\mathcal{M}$ 
\begin{equation}
  \hat{P}_\mathcal{M}(\bx) = \sum_{k\in\mathcal{M}} p_k G(\mathbf{x}|\bmu_k,\mathbf{\Sigma}_k) /\left(\zeta+ \hat{P}(\bx)\right)
  \label{eqn:gmm_macro_partitioning}.
\end{equation} 
A demonstration of this non-Gaussian fingerprint is also depicted in Fig.~\ref{fig:interlocked_circles}d.

\section{Probabilistic Motif Identifiers}

The main use case for PAMM involves the data-driven identification of local environments that characterize the state of atoms or molecular fragments, e.g. the secondary structure in a biomolecule or the crystal structure of a solid. 
For this reason, we will first demonstrate our approach by determining such probabilistic motif identifiers (PMIs) in two very different cases:  a Lennard-Jones cluster, that is an archetypical example of the problems encountered in nanomaterials, followed by the example of a 16 residue long hairpin peptide (GB1), which we discuss as a simple but realistic example of the applications to biological systems.

\subsection{Local Motifs of a Lennard-Jones Cluster}

The LJ$_{38}$ cluster has been used very often as a benchmark of minimization algorithms, free-energy techniques, and structure-recognition methods, because its potential energy landscape contains a very deep, narrow enthalpic minimum corresponding to a truncated \emph{fcc} lattice, and a broad basin containing a multitude of defective structures corresponding to icosahedral symmetry~\cite{walesdoye97jpca,wales98nature,lj38nested,ceri+10jctc}.
Here we used data from the $T=0.18T^\star$ replica of a long parallel tempering trajectory~\cite{ceri+10jctc}, that contains structures that are representative of both the solid phases and liquid-like, highly defective configurations.

Several of the possible environments (e.g. corner, edge, facet and core atoms) can be roughly identified by the number of nearest neighbors, that can be characterized based on purely radial information. Here we use the coordination number $c(i)$ for particle $i$, that we define as 
\begin{equation}
    c(i) = \sum_{j \ne i}^N \frac{1}{\exp\left(\frac{r_{ij}-r_\mathrm{c}}{\gamma}\right) + 1}
\end{equation}
where $\mathbf{r}_{ij}$ indicates the vector between particle $i$ and $j$ and $r_{ij}=|\mathbf{r}_{ij}|$ is the Euclidean distance between those two atoms. In order for the Fermi function to count the number of first neighbors, we set 
$r_\text{c}=$ 1.45$\sigma$ and $\gamma=$0.2$\sigma$.

\begin{figure*}[bth]
    \centering
    \includegraphics[width=0.95\linewidth]{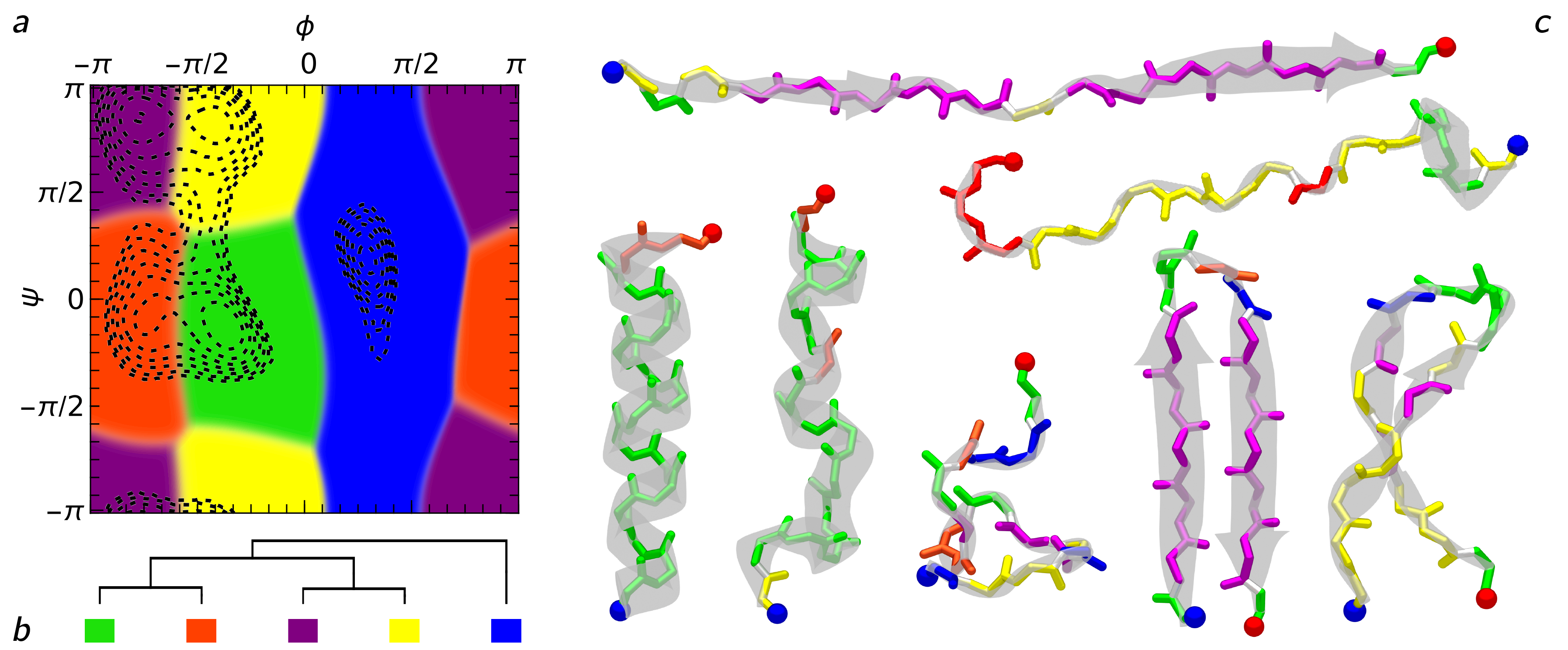}
\caption{
(a) PMIs for the backbone dihedrals of a $\beta$-hairpin are shown as a function of the Ramachandran angles. Different regions are colored based on the cluster that is associated with the dominant PMI. The underlying probability distribution is shown as black contours, equally spaced on a logarithmic scale. (b) Dendrogram representation of the hierarchical clustering of the adjacency matrix of the PAMM clusters. (c) A few representative  structures of the molecules are shown where the residues are colored according to the dominant PMI. The color code corresponds to that used in panel (a).}
\label{fig:bhairp_pmi}
\end{figure*}

While neighbor counts can be very effective in identifying motifs in the minimum-energy structures, finite-temperature simulations can be considerably fuzzier.
Furthermore, the coordination number cannot distinguish between bulk environments from a \emph{fcc} lattice and the core of an icosahedral motif. 
To address these problems, and to demonstrate the behavior of PAMM in a non-trivial example, we supplemented $c(i)$ with angular information.
We used the local bond order parameters (or Steinhardt order parameters) \cite{stei+83prb} $q_4(i)$ and $q_6(i)$, which are known to be able to distinguish body center cubic (\emph{bcc}), face centered cubic (\emph{fcc}) or hexagonal close packed (\emph{hcp}) crystal lattices \cite{moro+05prl,coas+07prb,ogat+92pra}.

We use the definition 
\begin{equation}
    q_{l}(i) = \sqrt{\frac{4\pi}{2l+1}\sum_{m=-l}^{l}|q_{lm}(i)|^2}
\end{equation}
in which the complex vector $q_{lm}(i)$ is defined as
\begin{equation}
    q_{lm}(i) = \frac{1}{c(i)}\sum_{j\ne i}^{N} \frac{Y_{lm}(\mathbf{r}_{ij})}{\exp\left(\frac{r_{ij}-r_\mathrm{c}}{\gamma}\right) + 1}
\end{equation}
The functions $Y_{lm}(\mathbf{r}_{ij})$ are the spherical harmonics and the loop runs over all particles, since the Fermi function singles out contributions from just the first coordination shell.
Since $c(i)$, $q_4(i)$ and $q_6(i)$ have very different ranges, when combining the different order parameters to give a 3D descriptor of the environments, we centered and scaled them so that each component has unit variance.

Figure~\ref{fig:lj38-env} demonstrates the outcome of a PAMM analysis for the finite-temperature trajectory.
A point based approach with a smoothing parameter $f_\mathrm{points}=0.02$ was used to compute the KDE, while for the clustering step the scaling factor $\alpha$ was set to $1$.
PAMM can recognize motifs based on both coordination number and angular correlations based on the three-dimensional joint probability density (Fig.~\ref{fig:lj38-env}a) -- identifying several more clusters than it would be possible based on $c(i)$ alone (Fig.~\ref{fig:lj38-env}c).
The PMIs for the different motifs (Fig.~\ref{fig:lj38-env}b) can be used to recognize the motifs in each snapshot of the simulation (Fig.~\ref{fig:lj38-env}d), and could be used, e.g., to bias a molecular dynamics simulation which triggers transitions between different structures. 
In this relatively simple case clusters are clear-cut, robust to sub-sampling and to variations of the PAMM parameters, and with an approximately Gaussian shape. Even though an agglomerative meta-clustering step is therefore not necessary, we did compute the cluster stability matrix, and generated the associated hierarchical clustering tree (Fig.~\ref{fig:lj38-env}e). 
Inspection of the binary tree is insightful, showing that individual clusters can be grouped in two main branches, corresponding to surface and bulk atoms.
As we will also see in other cases, it appears that the hierarchical merging of the PAMM clusters can be interpreted much in the same way as a disconnectivity graph~\cite{beck-karp97jcp,mort+02jcp}, reporting on the relations between different basins in pattern space.  

\subsection{Local Motifs of a $\beta$-hairpine Peptide}

Secondary structure in proteins is a textbook example of how molecular motifs can behave as building blocks of complex supramolecular structures.
Many algorithms exist to identify secondary structure patterns (SSPs), that are based on the identification of hydrogen bonds~\cite{fris+95prot,kabs+83bp}. 
More recently, it has been shown that secondary structure conformations can be classified solely on the basis of backbone dihedrals of a protein \cite{hollingsworth+12jmb,nagy+14jcim}.
As a demonstration of our classification approach to realistic periodic data, we have applied our method to analyze the data of the backbone dihedral pairs $(\phi,\psi)$ from a replica exchange simulation of a 16-residue C-terminal fragment of the immunoglobulin binding domain B1 of the Streptococcus protein G in explicit solvent (GB1, amino acids sequence
Ace-GEWTYDDATKTFTVTE-Nme) (for further details of the simulation see \citet{arde+15jctc}).
For each residue and each frame of the trajectory we computed the backbone dihedral angles $\phi$ and $\psi$, and performed a PAMM analysis. 
The underlying KDE was estimated by using a point-based KDE smoothing $f_\text{p}=0.15$, and step-scaling $\alpha=1.0$ for the subsequent Quick-Step clustering.
The resulting PMIs result in a partitioning of the Ramachandran $\phi-\psi$ plot~\cite{ramachandran1963stereochemistry} in 5 regions, that correspond roughly to $\beta$ sheets, $\alpha$ helices, turns, etc. (see Fig. \ref{fig:bhairp_pmi}), and that are clearly associated with local probability maxima in the KDE (shown as black contours in Fig. \ref{fig:bhairp_pmi}a). 
Even though the clusters are identified as von Mises modes, with a single basis function assigned to each of them, it is clear that the PMI correspond very accurately to the partitioning of the probability density in basins of attraction, with the transition zones between two PMIs following closely the dividing surface between basins.
In Fig. \ref{fig:bhairp_pmi}b we also show a few reference structures selected from the trajectory. The aminoacids in the backbone have been colored based on the dominant PMI to which they are associated.

It is easy to recognize the PMIs associated with well known secondary structure elements by comparing reference structures to the PAMM partitioning of the Ramachandran plot. One can clearly identify, e.g. $\alpha$-helices or the antiparallel $\beta$-sheets that are abundant in the simulation data for the GB1 fragment.
Chains of dihedrals that correspond to a $\beta$-sheet conformation are also seen in extended structures. We also find several instance of turn-type T1 motifs~\cite{nagy+14jcim}, that in our case concentrate in the unstructured portions of the polypeptide.

Meanwhile, it is clear that there is not a 1:1 correspondence between PMIs and ``traditional'' secondary structure motifs.  Some of the PMIs correspond to portions of the Ramachandran plot that are traditionally assigned to polyproline II (PPII) helices and left-handed helices, yet we could not identify a significant presence of stable structures associated with those motifs. Rather, extended structures and distorted $\beta$ strands are associated with the PPII region, while left-handed helical patterns appear, together with many other PMIs, within disordered, ``random-coil'' configurations.
This observation highlights the potential of a data-driven approach to the definition of molecular patterns. Well-separated clusters in fingerprint space are recognized even though they do not appear clearly by just inspecting the trajectory in search for well-established structural motifs. This agnostic behavior could be very useful, for instance, in rationalizing the behavior of intrinsically-disordered proteins~\cite{tompa+14molcell,rece+12cpps}, or to study polypeptides in unusual environments, such as at inorganic interfaces or in combination with synthetic polymers.
Furthermore, an automated probability analysis allows one to extend the definition of pattern space by combining e.g. several backbone dihedrals, or by combining dihedrals and H-bonding indicators. This procedure could give rise to more precise identification of secondary-structure patterns, and will be the subject of future research. 
Finally, the hierarchical clustering of the five PAMM motifs (Fig. \ref{fig:bhairp_pmi}b) shows another example of how the adjacency matrix built by a bootstrapping analysis reflects the structure of the free-energy landscape of patterns, with the linkage distance corresponding roughly to the free-energy barrier between basins.

\section{Global Structure Classification}
PAMM clustering of low-dimensional fingerprint spaces allowed us to demonstrate some of the improved features of this implementation, including periodic descriptors, adjacency-matrix determination and hierarchical ``meta-clustering''. 
We now move on to demonstrate the applicability of our method to problems with higher dimensionality. Rather than using higher-dimensional descriptors of local motifs, we consider the extreme case of attempting a global classification of the meta-stable states of the cluster and the polypeptide that we discussed in the previous Section. These problems are intrinsically high-dimensional, and do not simply involve an overdetermined description of a relatively low-dimensional problem. 
In order to simplify the interpretation of our results, we combined PAMM clustering with a two-dimensional representation of the configurational landscape. \mc{Among the many non-linear dimensionality reduction algorithms available~\cite{krusk64mds,joll86pca,abdi-lynn10pca,coif05pnas,maat08JMLR}, we used sketch-map~\cite{ceri+11pnas,trib+12pnas}, a method that was already used successfully to study these systems~\cite{ceri+13jctc,arde+15jctc}.} We  show that the combination of the two methods helps greatly overcoming some of their limitations, and understanding the relation between different structures in complex systems.

\begin{figure*}[t]
    \centering
    \includegraphics[width=\linewidth]{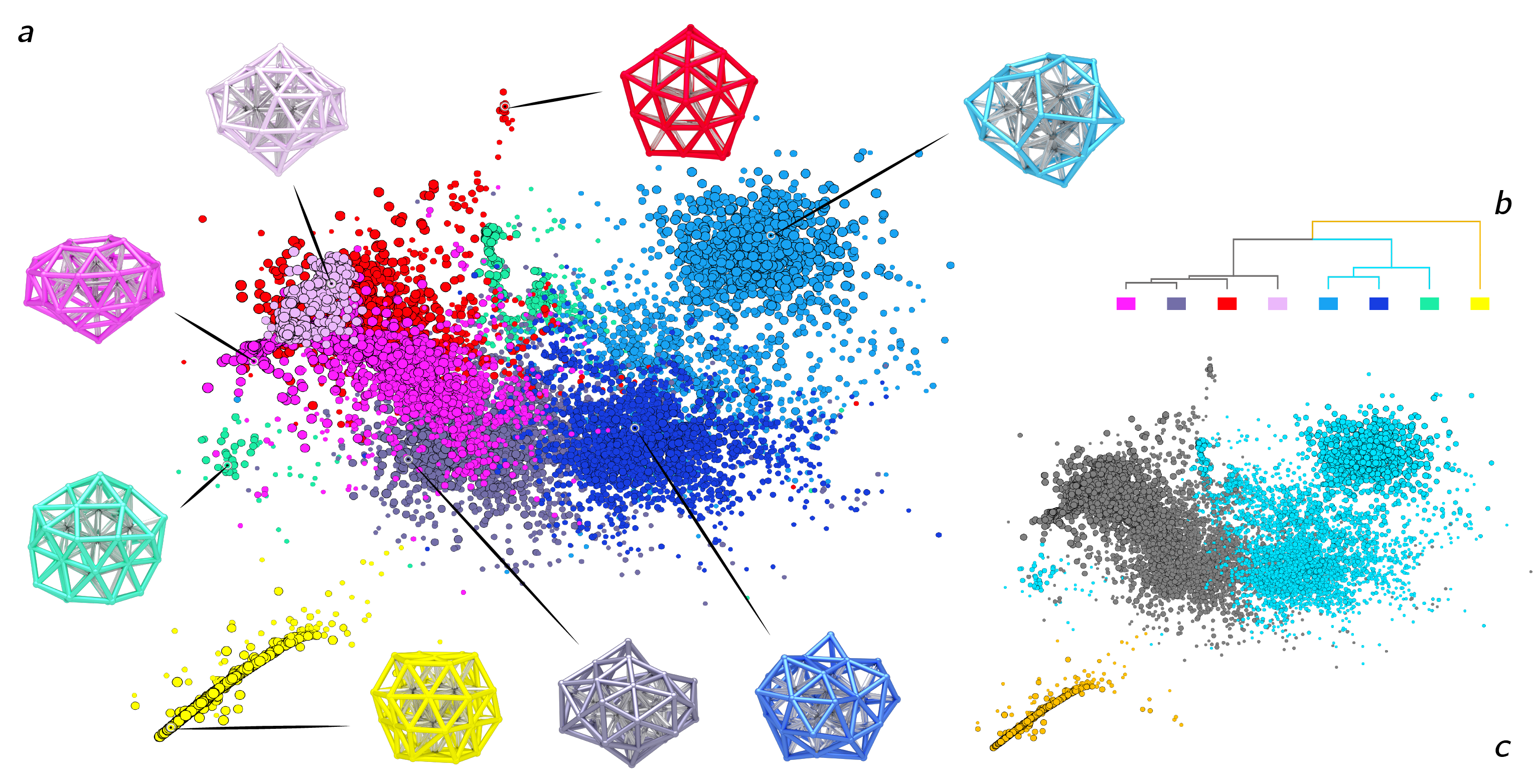}
\caption{PAMM classification of configurations extracted from a simulation of LJ$_{38}$ at $T^\star=0.18$. (a) The clusters generated from the first stage of PAMM using the parameters $f_\text{p}=0.05$ for the (point-based) KDE smoothing and $\alpha=1$ for the scaling of the Quick-Shift cutoff. Configurations from the trajectory are represented using a two-dimensional sketch-map representation. The size of points reflects the probability density. Points are colored according to the cluster to which they belong.
 (b) The initial clusters are merged according to the hierarchical clustering procedure. 
 (c) Sketch-map representation colored according to the macro-clusters, together with a snapshot representative of the highest-probability region for each motif.  }
\label{fig:lj38}
\end{figure*}

\subsection{Structural classification for LJ$_{38}$}

LJ$_{38}$ is one of the classical examples of simple clusters exhibiting a major structural transition, between a truncated-octahedral structure and a defective icosahedron. Since we consider a temperature close to the melting point, also liquid-like configurations should be present in the trajectory. 

In order to identify different structures, we use the same 15-dimensional descriptors based on a smooth histogram of coordination numbers that were introduced in Ref.~\citenum{ceri+13jctc}, and use them both as the basis for a PAMM analysis and as the input for the construction of a sketch-map representation, for which we used the same parameters and reference map as in Ref.~\citenum{ceri+13jctc}. 

We decided not to tune the fingerprints that had already been used in the literature so as to test the stability of PAMM-based clustering when dealing with sub-optimal inputs. For instance, some of the histogram bins have near-zero variance, which would yield singular bandwidth matrices if we did not stabilize the estimator with the OAS.
Furthermore, the sharp cutoff function used in defining the coordination number histogram leads to the presence of a large number of basins, corresponding to minute differences in the structures. However, by using a point-based localization with $f_\text{p}=0.05$, we implicitly consider a probabilistic model in which each cluster contains at least 5\%{} of the data. As a result, the smearing parameter and the Quick-Shift threshold are large enough that many of these minute clusters coalesce, leaving just 8 motifs detected (Figure~\ref{fig:lj38}a). 

The number and type of clusters, however, depend rather sensitively on the PAMM parameters. 
If one inspects the structural features that are associated with the initial PAMM clusters (see Figure~\ref{fig:lj38}a) it becomes clear that PAMM clustering does identify portions of configuration space that are clearly distinct, corresponding e.g. to different defective ``gemstone'' structures, to the truncated octahedron, and to liquid-like structures with different kinds of surface geometry. 
A dendrogram representation of the hierarchical clustering based on the adjacency matrix (Figure~\ref{fig:lj38}b) shows that some of the clusters are very sensitive to statistical noise, and resulting in high adjacency. One can identify three very stable meta-clusters, that are represented in (Figure~\ref{fig:lj38}c). PAMM recognizes that the most prominent features of the free-energy landscape comprise liquid-like states, more or less disordered icosahedral fragments and -- even more clearly separated --the truncated octahedron.
Once again it is remarkable how the adjacency matrix, despite being constructed just as a heuristic indicator of cluster stability, reflects the connectivity of the free-energy landscape. Although this far we were not able to reveal a formal connection, this aspect will be the subject of further investigation.

\begin{figure*}[t]
    \centering
    \includegraphics[width=\linewidth]{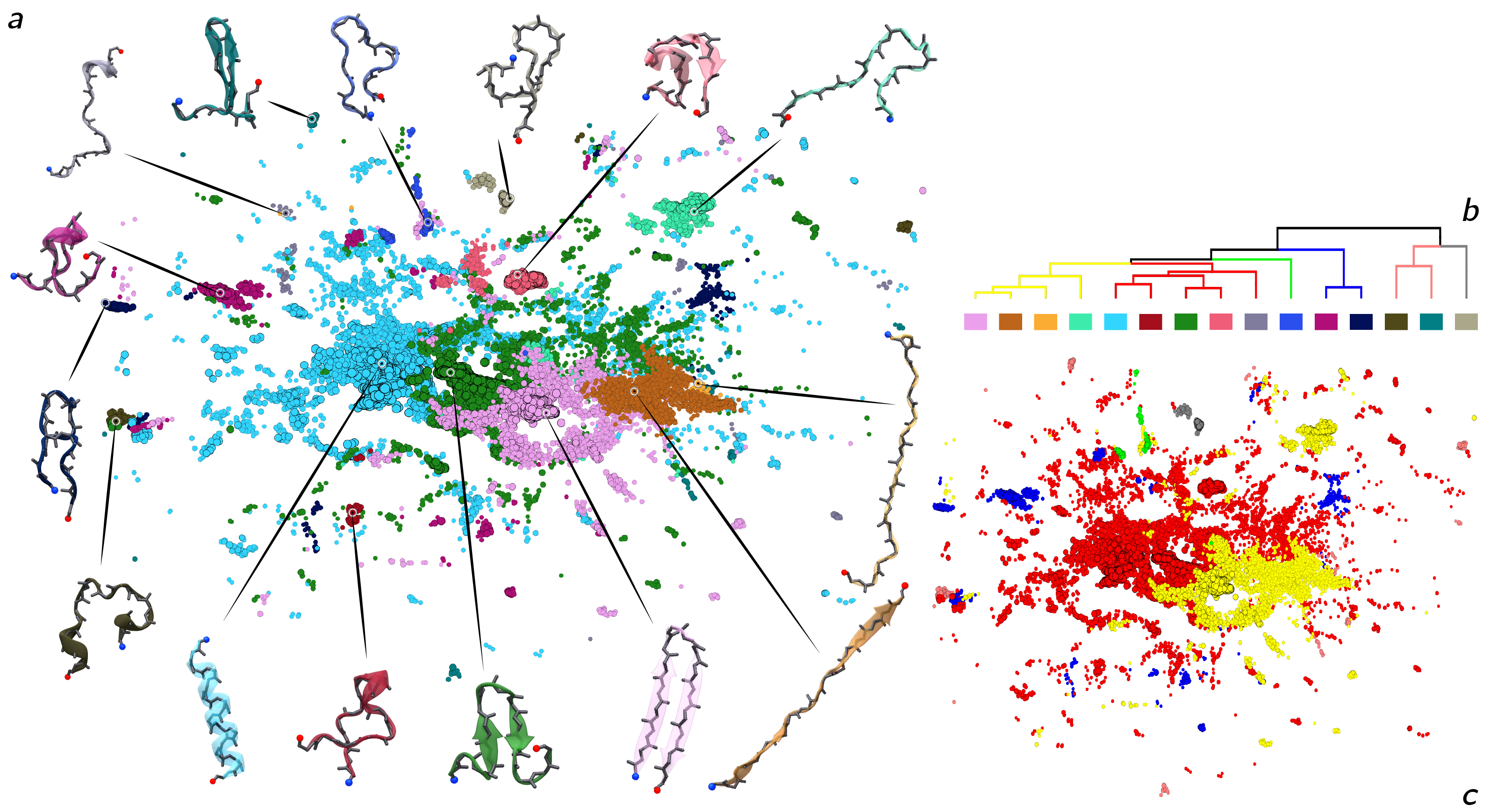}
\caption{PAMM classification of configurations extracted from a simulation of the GB1 hairpin fragment. (a) Clusters generated in the first stage of PAMM using the parameters $0.1$ for the (point-based) KDE smoothing and $\alpha=1$ for the scaling of the Quick-Shift cutoff. Configurations from the trajectory are represented using a two-dimensional sketch-map representation. The size of points reflects the probability density. Points are colored according to the cluster to which they belong.
 (b) The initial clusters are merged according to the hierarchical clustering procedure. 
 (c) Sketch-map representation colored according to the macro-clusters, together with a snapshot representative of the highest-probability region for each conformation.  }
\label{fig:gb1}
\end{figure*}

One final consideration concerns the relationship between the clustering and the dimensionality reduction approaches to describe complex atomistic systems.
A sketch-map representation does offer a more intuitive, bird's eye view of the landscape. 
However, in order to achieve such a high level of coarse graining, the dimensionality reduction algorithm introduces considerable distortion, including also discontinuities in the projection~\cite{trib+12pnas,ceri+13jctc,comi+17jcp}. 
For this reason, it is important to perform the clustering step in the high-dimensional space to avoid having a classification which is biased by sketch-map artifacts.
For instance, Figure~\ref{fig:lj38} shows clearly that one of the clusters identified by PAMM is split in two by the projection, without a continuous path connecting the fragments on the map. 
Thus, combining clustering and dimensionality-reduction approaches might provide a strategy to compensate for the shortcomings of the two methods, and obtain deeper understanding of the structural features of the system.

\subsection{Structural classification for a  $\beta$-hairpine Peptide}

The structural landscape for the GB1 oligopeptide provides another suitable benchmark for the application of PAMM to the clustering of high-dimensional data. This $\beta$-hairpin fragment has been studied extensively by metadynamics~\cite{bono+08jacs}, and has been used to demonstrate the comparison between wild type and mutant proteins using sketch-map~\cite{arde+15jctc}. Such analysis revealed a rugged free-energy landscape, containing many metastable states including a helical configuration, several mis-folded hairpin configurations as well as the native fold. 

As in the case of the the LJ$_{38}$ cluster, we used a simple-minded choice of high-dimensional descriptors -- namely, the 30 backbone dihedrals -- as the input representation. This choice minimizes the bias on the clustering procedure, but implies that configurations that differ by minor details (e.g. the configuration of the terminal aminoacids) are considered as distinct structures.  
Indeed, a global classification of the GB1 peptide produces a multitude of clusters (Figure~\ref{fig:gb1}a), and varies considerably depending on the clustering details and bootstrapping resampling. Hierarchical cluster merging, however,  simplifies considerably this picture (Figure~\ref{fig:gb1}b), making it possible to group states that are associated with a helical configuration, the native fold, and a few misfolded hairpin configurations (Figure~\ref{fig:gb1}c).
Clustering also highlights the limitations of the sketch-map projection, that gives a contiguous representation of the high-probability clusters (helix and native hairpin) but scatters higher-free-energy states at the periphery of the map. 

\section{Conclusions}
We presented a density-based clustering approach to analyze the outcome of atomistic simulations in search for the essential atomic-scale patterns that characterize the behavior of materials and molecules. 
The algorithm generalizes the recently-introduced PAMM method to periodic, high-dimensional and/or sparsely sampled data, combining several state-of-the-art techniques to guarantee robust, reliable and efficient clustering. 
Crucially, the method we introduce performs repeated clustering attempts on top of re-sampled distributions constructed by bootstrapping, and uses these to construct an adjacency matrix that characterizes the stability and the overlaps between clusters. 
Combined with a hierarchical ``meta-clustering'' and a binary tree representation this further analysis serves two purposes. 
First, it provides a way to improve the quality of PAMM clusters -- by merging non clear-cut motifs and allowing the representation of strongly non-Gaussian modes in the probability.
Second, it makes it possible to recognize the relations between the fine-grained details of the free-energy landscape and the main basins on a coarser scale.  In all the examples we considered, the hierarchical clustering reflects qualitatively the structure of the free energy, resembling a disconnectivity graph. It might be possible to formulate the construction of the adjacency matrix in a way that makes this analogy quantitative, an endeavor which will be the subject of future research. 

We demonstrate that this improved PAMM algorithm identifies coordination environments in a LJ$_{38}$ cluster and secondary-structure motifs in a 16-residues $\beta$-hairpin  peptide. Although our method is geared towards recognizing \emph{local} molecular patterns \mc{ -- and more principled approaches exist that are designed to identify \emph{global} conformational states~\cite{webe+04pcca,deuf-webw05pcca,robl-webe13pcca+,chod-noe2014pcca} --} we also show it can cluster overall structures in these two systems. 
The combination of PAMM clustering, binary-tree merging of the motifs and sketch-map representation of high-dimensional free energy landscapes provide multiple insights into the behavior of complex atomistic systems, overcoming some of the limitations of the methods, such as the presence of discontinuities in sketch-map projections. 
Development code to perform a PAMM analysis is available from an on-line repository~\footnote{{http://cosmo-epfl.github.io}}, and the data we used for demonstration purposes is provided in the SI.

The probabilistic motif identifiers associated with each cluster can be used as fuzzy, smoothly varying collective variables in biased molecular dynamics schemes to accelerate sampling and reconstruct the underlying free-energy landscape. The Gaussian Mixture Model associated with the PMI could also be used as an ansatz for the target probability distribution in a variational-sampling scheme~\cite{vals-parr14prl}, with the populations of the clusters as the variational parameters. 
Data-driven approaches to recognize the building blocks of complex materials constitute a necessary ingredient to assist the analysis and interpretation of large-scale atomistic trajectories, and provide a natural representation of free-energy landscapes in terms of the most stable configurations that can be used to accelerate configurational sampling.

\begin{acknowledgement}
The Authors would like to thank Gareth Tribello for insightful discussion and help implementing PMIs in PLUMED. PG acknowledges funding from the MPG-EPFL center for molecular nanoscience, and from the NCCR MARVEL, funded by the Swiss National Science Foundation.. 
MC and RHM were supported by the European Research Council under the European Union's Horizon 2020 research and innovation programme (grant agreement no. 677013-HBMAP). 
\end{acknowledgement}
\providecommand{\latin}[1]{#1}
\providecommand*\mcitethebibliography{\thebibliography}
\csname @ifundefined\endcsname{endmcitethebibliography}
  {\let\endmcitethebibliography\endthebibliography}{}


\begin{mcitethebibliography}{92}
\providecommand*\natexlab[1]{#1}
\providecommand*\mciteSetBstSublistMode[1]{}
\providecommand*\mciteSetBstMaxWidthForm[2]{}
\providecommand*\mciteBstWouldAddEndPuncttrue
  {\def\EndOfBibitem{\unskip.}}
\providecommand*\mciteBstWouldAddEndPunctfalse
  {\let\EndOfBibitem\relax}
\providecommand*\mciteSetBstMidEndSepPunct[3]{}
\providecommand*\mciteSetBstSublistLabelBeginEnd[3]{}
\providecommand*\EndOfBibitem{}
\mciteSetBstSublistMode{f}
\mciteSetBstMaxWidthForm{subitem}{(\alph{mcitesubitemcount})}
\mciteSetBstSublistLabelBeginEnd
  {\mcitemaxwidthsubitemform\space}
  {\relax}
  {\relax}

\bibitem[Fischer \latin{et~al.}(2006)Fischer, Tibbetts, Morgan, and
  Ceder]{fisc2006natmat}
Fischer,~C.~C.; Tibbetts,~K.~J.; Morgan,~D.; Ceder,~G. Predicting crystal
  structure by merging data mining with quantum mechanics. \emph{Nature
  materials} \textbf{2006}, \emph{5}, 641\relax
\mciteBstWouldAddEndPuncttrue
\mciteSetBstMidEndSepPunct{\mcitedefaultmidpunct}
{\mcitedefaultendpunct}{\mcitedefaultseppunct}\relax
\EndOfBibitem
\bibitem[Marzari(2016)]{marzari2016materials}
Marzari,~N. Materials modelling: The frontiers and the challenges. \emph{Nature
  materials} \textbf{2016}, \emph{15}, 381--382\relax
\mciteBstWouldAddEndPuncttrue
\mciteSetBstMidEndSepPunct{\mcitedefaultmidpunct}
{\mcitedefaultendpunct}{\mcitedefaultseppunct}\relax
\EndOfBibitem
\bibitem[Lindorff-Larsen \latin{et~al.}(2016)Lindorff-Larsen, Maragakis, Piana,
  and Shaw]{lind+16jpcb}
Lindorff-Larsen,~K.; Maragakis,~P.; Piana,~S.; Shaw,~D.~E. Picosecond to
  Millisecond Structural Dynamics in Human Ubiquitin. \emph{The Journal of
  Physical Chemistry B} \textbf{2016}, \emph{120}, 8313--8320\relax
\mciteBstWouldAddEndPuncttrue
\mciteSetBstMidEndSepPunct{\mcitedefaultmidpunct}
{\mcitedefaultendpunct}{\mcitedefaultseppunct}\relax
\EndOfBibitem
\bibitem[Zhao \latin{et~al.}(2013)Zhao, Perilla, Yufenyuy, Meng, Chen, Ning,
  Ahn, Gronenborn, Schulten, Aiken, and Zhang]{zhao+13nat}
Zhao,~G.; Perilla,~J.~R.; Yufenyuy,~E.~L.; Meng,~X.; Chen,~B.; Ning,~J.;
  Ahn,~J.; Gronenborn,~A.~M.; Schulten,~K.; Aiken,~C.; Zhang,~P. Mature HIV-1
  capsid structure by cryo-electron microscopy and all-atom molecular dynamics.
  \emph{Nature} \textbf{2013}, \emph{497}, 643--646\relax
\mciteBstWouldAddEndPuncttrue
\mciteSetBstMidEndSepPunct{\mcitedefaultmidpunct}
{\mcitedefaultendpunct}{\mcitedefaultseppunct}\relax
\EndOfBibitem
\bibitem[Agrawal and Choudhary(2016)Agrawal, and Choudhary]{agra16apl}
Agrawal,~A.; Choudhary,~A. Perspective: Materials informatics and big data:
  Realization of the “fourth paradigm” of science in materials science.
  \emph{APL Materials} \textbf{2016}, \emph{4}, 053208\relax
\mciteBstWouldAddEndPuncttrue
\mciteSetBstMidEndSepPunct{\mcitedefaultmidpunct}
{\mcitedefaultendpunct}{\mcitedefaultseppunct}\relax
\EndOfBibitem
\bibitem[Bereau \latin{et~al.}(2016)Bereau, Andrienko, and Kremer]{bere16aplm}
Bereau,~T.; Andrienko,~D.; Kremer,~K. Research Update: Computational materials
  discovery in soft matter. \emph{APL Materials} \textbf{2016}, \emph{4},
  053101\relax
\mciteBstWouldAddEndPuncttrue
\mciteSetBstMidEndSepPunct{\mcitedefaultmidpunct}
{\mcitedefaultendpunct}{\mcitedefaultseppunct}\relax
\EndOfBibitem
\bibitem[Curtarolo \latin{et~al.}(2013)Curtarolo, Hart, Nardelli, Mingo,
  Sanvito, and Levy]{curt+13nat}
Curtarolo,~S.; Hart,~G.~L.; Nardelli,~M.~B.; Mingo,~N.; Sanvito,~S.; Levy,~O.
  The high-throughput highway to computational materials design. \emph{Nature
  materials} \textbf{2013}, \emph{12}, 191\relax
\mciteBstWouldAddEndPuncttrue
\mciteSetBstMidEndSepPunct{\mcitedefaultmidpunct}
{\mcitedefaultendpunct}{\mcitedefaultseppunct}\relax
\EndOfBibitem
\bibitem[Phillips and Littlewood(2016)Phillips, and Littlewood]{phil16aplm}
Phillips,~C.~L.; Littlewood,~P. Preface: Special Topic on Materials Genome.
  \emph{APL Materials} \textbf{2016}, \emph{4}, 053001\relax
\mciteBstWouldAddEndPuncttrue
\mciteSetBstMidEndSepPunct{\mcitedefaultmidpunct}
{\mcitedefaultendpunct}{\mcitedefaultseppunct}\relax
\EndOfBibitem
\bibitem[Pizzi \latin{et~al.}(2016)Pizzi, Cepellotti, Sabatini, Marzari, and
  Kozinsky]{pizzi2016}
Pizzi,~G.; Cepellotti,~A.; Sabatini,~R.; Marzari,~N.; Kozinsky,~B. {AiiDA}:
  automated interactive infrastructure and database for computational science.
  \emph{Computational Materials Science} \textbf{2016}, \emph{111},
  218--230\relax
\mciteBstWouldAddEndPuncttrue
\mciteSetBstMidEndSepPunct{\mcitedefaultmidpunct}
{\mcitedefaultendpunct}{\mcitedefaultseppunct}\relax
\EndOfBibitem
\bibitem[Chodera \latin{et~al.}(2007)Chodera, Singhal, Pande, Dill, and
  Swope]{chod07jcp}
Chodera,~J.~D.; Singhal,~N.; Pande,~V.~S.; Dill,~K.~A.; Swope,~W.~C. Automatic
  discovery of metastable states for the construction of Markov models of
  macromolecular conformational dynamics. \emph{The Journal of chemical
  physics} \textbf{2007}, \emph{126}, 04B616\relax
\mciteBstWouldAddEndPuncttrue
\mciteSetBstMidEndSepPunct{\mcitedefaultmidpunct}
{\mcitedefaultendpunct}{\mcitedefaultseppunct}\relax
\EndOfBibitem
\bibitem[de~Groot \latin{et~al.}(2001)de~Groot, Daura, Mark, and
  Grubm{\"u}ller]{de01jmb}
de~Groot,~B.~L.; Daura,~X.; Mark,~A.~E.; Grubm{\"u}ller,~H. Essential dynamics
  of reversible peptide folding: memory-free conformational dynamics governed
  by internal hydrogen bonds. \emph{Journal of molecular biology}
  \textbf{2001}, \emph{309}, 299--313\relax
\mciteBstWouldAddEndPuncttrue
\mciteSetBstMidEndSepPunct{\mcitedefaultmidpunct}
{\mcitedefaultendpunct}{\mcitedefaultseppunct}\relax
\EndOfBibitem
\bibitem[Andrec \latin{et~al.}(2005)Andrec, Felts, Gallicchio, and
  Levy]{andr05pnas}
Andrec,~M.; Felts,~A.~K.; Gallicchio,~E.; Levy,~R.~M. Protein folding pathways
  from replica exchange simulations and a kinetic network model.
  \emph{Proceedings of the National Academy of Sciences of the United States of
  America} \textbf{2005}, \emph{102}, 6801--6806\relax
\mciteBstWouldAddEndPuncttrue
\mciteSetBstMidEndSepPunct{\mcitedefaultmidpunct}
{\mcitedefaultendpunct}{\mcitedefaultseppunct}\relax
\EndOfBibitem
\bibitem[Singhal \latin{et~al.}(2004)Singhal, Snow, and Pande]{sing04jcp}
Singhal,~N.; Snow,~C.~D.; Pande,~V.~S. Using path sampling to build better
  Markovian state models: predicting the folding rate and mechanism of a
  tryptophan zipper beta hairpin. \emph{The Journal of chemical physics}
  \textbf{2004}, \emph{121}, 415--425\relax
\mciteBstWouldAddEndPuncttrue
\mciteSetBstMidEndSepPunct{\mcitedefaultmidpunct}
{\mcitedefaultendpunct}{\mcitedefaultseppunct}\relax
\EndOfBibitem
\bibitem[Huan \latin{et~al.}(2015)Huan, Mannodi-Kanakkithodi, and
  Ramprasad]{huan+15prb}
Huan,~T.~D.; Mannodi-Kanakkithodi,~A.; Ramprasad,~R. Accelerated materials
  property predictions and design using motif-based fingerprints.
  \emph{Physical Review B} \textbf{2015}, \emph{92}, 014106\relax
\mciteBstWouldAddEndPuncttrue
\mciteSetBstMidEndSepPunct{\mcitedefaultmidpunct}
{\mcitedefaultendpunct}{\mcitedefaultseppunct}\relax
\EndOfBibitem
\bibitem[Rupp \latin{et~al.}(2012)Rupp, Tkatchenko, M{\"u}ller, and
  Von~Lilienfeld]{rupp12prl}
Rupp,~M.; Tkatchenko,~A.; M{\"u}ller,~K.-R.; Von~Lilienfeld,~O.~A. Fast and
  accurate modeling of molecular atomization energies with machine learning.
  \emph{Physical review letters} \textbf{2012}, \emph{108}, 058301\relax
\mciteBstWouldAddEndPuncttrue
\mciteSetBstMidEndSepPunct{\mcitedefaultmidpunct}
{\mcitedefaultendpunct}{\mcitedefaultseppunct}\relax
\EndOfBibitem
\bibitem[Faber \latin{et~al.}(2016)Faber, Lindmaa, von Lilienfeld, and
  Armiento]{fabe16prl}
Faber,~F.~A.; Lindmaa,~A.; von Lilienfeld,~O.~A.; Armiento,~R. Machine Learning
  Energies of 2 Million Elpasolite (A B C 2 D 6) Crystals. \emph{Physical
  Review Letters} \textbf{2016}, \emph{117}, 135502\relax
\mciteBstWouldAddEndPuncttrue
\mciteSetBstMidEndSepPunct{\mcitedefaultmidpunct}
{\mcitedefaultendpunct}{\mcitedefaultseppunct}\relax
\EndOfBibitem
\bibitem[Behler(2016)]{behl16jcp}
Behler,~J. Perspective: Machine learning potentials for atomistic simulations.
  \emph{The Journal of Chemical Physics} \textbf{2016}, \emph{145},
  170901\relax
\mciteBstWouldAddEndPuncttrue
\mciteSetBstMidEndSepPunct{\mcitedefaultmidpunct}
{\mcitedefaultendpunct}{\mcitedefaultseppunct}\relax
\EndOfBibitem
\bibitem[De \latin{et~al.}(2017)De, Musil, Ingram, Baldauf, and
  Ceriotti]{de17jchmi}
De,~S.; Musil,~F.; Ingram,~T.; Baldauf,~C.; Ceriotti,~M. Mapping and
  classifying molecules from a high-throughput structural database.
  \emph{Journal of Cheminformatics} \textbf{2017}, \emph{9}, 6\relax
\mciteBstWouldAddEndPuncttrue
\mciteSetBstMidEndSepPunct{\mcitedefaultmidpunct}
{\mcitedefaultendpunct}{\mcitedefaultseppunct}\relax
\EndOfBibitem
\bibitem[Sparks \latin{et~al.}(2016)Sparks, Gaultois, Oliynyk, Brgoch, and
  Meredig]{spar16scrm}
Sparks,~T.~D.; Gaultois,~M.~W.; Oliynyk,~A.; Brgoch,~J.; Meredig,~B. Data
  mining our way to the next generation of thermoelectrics. \emph{Scripta
  Materialia} \textbf{2016}, \emph{111}, 10--15\relax
\mciteBstWouldAddEndPuncttrue
\mciteSetBstMidEndSepPunct{\mcitedefaultmidpunct}
{\mcitedefaultendpunct}{\mcitedefaultseppunct}\relax
\EndOfBibitem
\bibitem[Rost and Sander(1994)Rost, and Sander]{rost+94prot}
Rost,~B.; Sander,~C. Combining evolutionary information and neural networks to
  predict protein secondary structure. \emph{Proteins: Structure, Function, and
  Bioinformatics} \textbf{1994}, \emph{19}, 55--72\relax
\mciteBstWouldAddEndPuncttrue
\mciteSetBstMidEndSepPunct{\mcitedefaultmidpunct}
{\mcitedefaultendpunct}{\mcitedefaultseppunct}\relax
\EndOfBibitem
\bibitem[Ballester and Mitchell(2010)Ballester, and Mitchell]{ball2010mach}
Ballester,~P.~J.; Mitchell,~J.~B. A machine learning approach to predicting
  protein--ligand binding affinity with applications to molecular docking.
  \emph{Bioinformatics} \textbf{2010}, \emph{26}, 1169--1175\relax
\mciteBstWouldAddEndPuncttrue
\mciteSetBstMidEndSepPunct{\mcitedefaultmidpunct}
{\mcitedefaultendpunct}{\mcitedefaultseppunct}\relax
\EndOfBibitem
\bibitem[Cheng and Baldi(2006)Cheng, and Baldi]{chen06bion}
Cheng,~J.; Baldi,~P. A machine learning information retrieval approach to
  protein fold recognition. \emph{Bioinformatics} \textbf{2006}, \emph{22},
  1456--1463\relax
\mciteBstWouldAddEndPuncttrue
\mciteSetBstMidEndSepPunct{\mcitedefaultmidpunct}
{\mcitedefaultendpunct}{\mcitedefaultseppunct}\relax
\EndOfBibitem
\bibitem[Dale \latin{et~al.}(2010)Dale, Popescu, and Karp]{dale10bioi}
Dale,~J.~M.; Popescu,~L.; Karp,~P.~D. Machine learning methods for metabolic
  pathway prediction. \emph{BMC bioinformatics} \textbf{2010}, \emph{11},
  15\relax
\mciteBstWouldAddEndPuncttrue
\mciteSetBstMidEndSepPunct{\mcitedefaultmidpunct}
{\mcitedefaultendpunct}{\mcitedefaultseppunct}\relax
\EndOfBibitem
\bibitem[Haranczyk and Sethian(2010)Haranczyk, and Sethian]{hara+10jctc}
Haranczyk,~M.; Sethian,~J.~A. Automatic structure analysis in high-throughput
  characterization of porous materials. \emph{Journal of chemical theory and
  computation} \textbf{2010}, \emph{6}, 3472--3480\relax
\mciteBstWouldAddEndPuncttrue
\mciteSetBstMidEndSepPunct{\mcitedefaultmidpunct}
{\mcitedefaultendpunct}{\mcitedefaultseppunct}\relax
\EndOfBibitem
\bibitem[Carr \latin{et~al.}(2009)Carr, Lach-hab, Yang, Vaisman, and
  Blaisten-Barojas]{carr09mmm}
Carr,~D.~A.; Lach-hab,~M.; Yang,~S.; Vaisman,~I.~I.; Blaisten-Barojas,~E.
  Machine learning approach for structure-based zeolite classification.
  \emph{Microporous and Mesoporous Materials} \textbf{2009}, \emph{117},
  339--349\relax
\mciteBstWouldAddEndPuncttrue
\mciteSetBstMidEndSepPunct{\mcitedefaultmidpunct}
{\mcitedefaultendpunct}{\mcitedefaultseppunct}\relax
\EndOfBibitem
\bibitem[Sch{\"u}tt \latin{et~al.}(2014)Sch{\"u}tt, Glawe, Brockherde, Sanna,
  M{\"u}ller, and Gross]{schu14prb}
Sch{\"u}tt,~K.; Glawe,~H.; Brockherde,~F.; Sanna,~A.; M{\"u}ller,~K.; Gross,~E.
  How to represent crystal structures for machine learning: Towards fast
  prediction of electronic properties. \emph{Physical Review B} \textbf{2014},
  \emph{89}, 205118\relax
\mciteBstWouldAddEndPuncttrue
\mciteSetBstMidEndSepPunct{\mcitedefaultmidpunct}
{\mcitedefaultendpunct}{\mcitedefaultseppunct}\relax
\EndOfBibitem
\bibitem[Kayala and Baldi(2012)Kayala, and Baldi]{kaya12jcim}
Kayala,~M.~A.; Baldi,~P. Reactionpredictor: Prediction of complex chemical
  reactions at the mechanistic level using machine learning. \emph{Journal of
  chemical information and modeling} \textbf{2012}, \emph{52}, 2526--2540\relax
\mciteBstWouldAddEndPuncttrue
\mciteSetBstMidEndSepPunct{\mcitedefaultmidpunct}
{\mcitedefaultendpunct}{\mcitedefaultseppunct}\relax
\EndOfBibitem
\bibitem[Gasparotto and Ceriotti(2014)Gasparotto, and Ceriotti]{gasp-ceri14jcp}
Gasparotto,~P.; Ceriotti,~M. {Recognizing molecular patterns by machine
  learning: An agnostic structural definition of the hydrogen bond}. \emph{J.
  Chem. Phys.} \textbf{2014}, \emph{141}, 174110\relax
\mciteBstWouldAddEndPuncttrue
\mciteSetBstMidEndSepPunct{\mcitedefaultmidpunct}
{\mcitedefaultendpunct}{\mcitedefaultseppunct}\relax
\EndOfBibitem
\bibitem[Gasparotto \latin{et~al.}(2016)Gasparotto, Hassanali, and
  Ceriotti]{gasp+16jctc}
Gasparotto,~P.; Hassanali,~A.~A.; Ceriotti,~M. {Probing Defects and
  Correlations in the Hydrogen-Bond Network of ab Initio Water}. \emph{J. Chem.
  Theory Comput.} \textbf{2016}, \emph{12}, 1953--1964\relax
\mciteBstWouldAddEndPuncttrue
\mciteSetBstMidEndSepPunct{\mcitedefaultmidpunct}
{\mcitedefaultendpunct}{\mcitedefaultseppunct}\relax
\EndOfBibitem
\bibitem[Rossi \latin{et~al.}(2016)Rossi, Ceriotti, and
  Manolopoulos]{ross16jpcl}
Rossi,~M.; Ceriotti,~M.; Manolopoulos,~D.~E. Nuclear Quantum Effects in H+ and
  OH--Diffusion along Confined Water Wires. \emph{The Journal of Physical
  Chemistry Letters} \textbf{2016}, \emph{7}, 3001--3007\relax
\mciteBstWouldAddEndPuncttrue
\mciteSetBstMidEndSepPunct{\mcitedefaultmidpunct}
{\mcitedefaultendpunct}{\mcitedefaultseppunct}\relax
\EndOfBibitem
\bibitem[Behler and Parrinello(2007)Behler, and Parrinello]{behl-parr07prl}
Behler,~J.; Parrinello,~M. {Generalized Neural-Network Representation of
  High-Dimensional Potential-Energy Surfaces}. \emph{Phys. Rev. Lett.}
  \textbf{2007}, \emph{98}, 146401\relax
\mciteBstWouldAddEndPuncttrue
\mciteSetBstMidEndSepPunct{\mcitedefaultmidpunct}
{\mcitedefaultendpunct}{\mcitedefaultseppunct}\relax
\EndOfBibitem
\bibitem[Bart\'{o}k \latin{et~al.}(2013)Bart\'{o}k, Gillan, Manby, and
  Cs\'{a}nyi]{bart+13prb}
Bart\'{o}k,~A.~P.; Gillan,~M.~J.; Manby,~F.~R.; Cs\'{a}nyi,~G.
  {Machine-learning approach for one- and two-body corrections to density
  functional theory: Applications to molecular and condensed water}.
  \emph{Phys. Rev. B} \textbf{2013}, \emph{88}, 054104\relax
\mciteBstWouldAddEndPuncttrue
\mciteSetBstMidEndSepPunct{\mcitedefaultmidpunct}
{\mcitedefaultendpunct}{\mcitedefaultseppunct}\relax
\EndOfBibitem
\bibitem[Sadeghi \latin{et~al.}(2013)Sadeghi, Ghasemi, Schaefer, Mohr, Lill,
  and Goedecker]{sade+13jcp}
Sadeghi,~A.; Ghasemi,~S.~A.; Schaefer,~B.; Mohr,~S.; Lill,~M.~A.; Goedecker,~S.
  {Metrics for measuring distances in configuration spaces.} \emph{J. Chem.
  Phys.} \textbf{2013}, \emph{139}, 184118\relax
\mciteBstWouldAddEndPuncttrue
\mciteSetBstMidEndSepPunct{\mcitedefaultmidpunct}
{\mcitedefaultendpunct}{\mcitedefaultseppunct}\relax
\EndOfBibitem
\bibitem[Behler(2011)]{behl11jcp}
Behler,~J. {Atom-centered symmetry functions for constructing high-dimensional
  neural network potentials}. \emph{J. Chem. Phys.} \textbf{2011},
  \emph{134}\relax
\mciteBstWouldAddEndPuncttrue
\mciteSetBstMidEndSepPunct{\mcitedefaultmidpunct}
{\mcitedefaultendpunct}{\mcitedefaultseppunct}\relax
\EndOfBibitem
\bibitem[Pietrucci and Andreoni(2011)Pietrucci, and Andreoni]{piet-andr11prl}
Pietrucci,~F.; Andreoni,~W. {Graph Theory Meets Ab Initio Molecular Dynamics:
  Atomic Structures and Transformations at the Nanoscale}. \emph{Phys. Rev.
  Lett.} \textbf{2011}, \emph{107}, 085504\relax
\mciteBstWouldAddEndPuncttrue
\mciteSetBstMidEndSepPunct{\mcitedefaultmidpunct}
{\mcitedefaultendpunct}{\mcitedefaultseppunct}\relax
\EndOfBibitem
\bibitem[Kitao and Go(1999)Kitao, and Go]{KITAO1999164}
Kitao,~A.; Go,~N. Investigating protein dynamics in collective coordinate
  space. \emph{Current Opinion in Structural Biology} \textbf{1999}, \emph{9},
  164 -- 169\relax
\mciteBstWouldAddEndPuncttrue
\mciteSetBstMidEndSepPunct{\mcitedefaultmidpunct}
{\mcitedefaultendpunct}{\mcitedefaultseppunct}\relax
\EndOfBibitem
\bibitem[Branduardi \latin{et~al.}(2007)Branduardi, Gervasio, and
  Parrinello]{branda2007}
Branduardi,~D.; Gervasio,~F.~L.; Parrinello,~M. From A to B in free energy
  space. \emph{The Journal of Chemical Physics} \textbf{2007}, \emph{126},
  054103\relax
\mciteBstWouldAddEndPuncttrue
\mciteSetBstMidEndSepPunct{\mcitedefaultmidpunct}
{\mcitedefaultendpunct}{\mcitedefaultseppunct}\relax
\EndOfBibitem
\bibitem[N{\"u}ske \latin{et~al.}(2014)N{\"u}ske, Keller,
  P{\'e}rez-Hern{\'a}ndez, Mey, and Noé]{nuske2014variational}
N{\"u}ske,~F.; Keller,~B.~G.; P{\'e}rez-Hern{\'a}ndez,~G.; Mey,~A.~S.;
  Noé,~F. Variational approach to molecular kinetics. \emph{Journal of
  chemical theory and computation} \textbf{2014}, \emph{10}, 1739--1752\relax
\mciteBstWouldAddEndPuncttrue
\mciteSetBstMidEndSepPunct{\mcitedefaultmidpunct}
{\mcitedefaultendpunct}{\mcitedefaultseppunct}\relax
\EndOfBibitem
\bibitem[Tiwary and Berne(2016)Tiwary, and Berne]{tiwary2016spectral}
Tiwary,~P.; Berne,~B. Spectral gap optimization of order parameters for
  sampling complex molecular systems. \emph{Proceedings of the National Academy
  of Sciences} \textbf{2016}, \emph{113}, 2839--2844\relax
\mciteBstWouldAddEndPuncttrue
\mciteSetBstMidEndSepPunct{\mcitedefaultmidpunct}
{\mcitedefaultendpunct}{\mcitedefaultseppunct}\relax
\EndOfBibitem
\bibitem[Rosenkrantz \latin{et~al.}(1977)Rosenkrantz, Stearns, and Philip
  M.~Lewis]{rosenkrantz+77siam}
Rosenkrantz,~D.~J.; Stearns,~R.~E.; Philip M.~Lewis,~I. An Analysis of Several
  Heuristics for the Traveling Salesman Problem. \emph{SIAM Journal on
  Computing} \textbf{1977}, \emph{6}, 563--581\relax
\mciteBstWouldAddEndPuncttrue
\mciteSetBstMidEndSepPunct{\mcitedefaultmidpunct}
{\mcitedefaultendpunct}{\mcitedefaultseppunct}\relax
\EndOfBibitem
\bibitem[Ceriotti \latin{et~al.}(2013)Ceriotti, Tribello, and
  Parrinello]{ceri+13jctc}
Ceriotti,~M.; Tribello,~G.~A.; Parrinello,~M. {Demonstrating the
  Transferability and the Descriptive Power of Sketch-Map}. \emph{J. Chem.
  Theory Comput.} \textbf{2013}, \emph{9}, 1521--1532\relax
\mciteBstWouldAddEndPuncttrue
\mciteSetBstMidEndSepPunct{\mcitedefaultmidpunct}
{\mcitedefaultendpunct}{\mcitedefaultseppunct}\relax
\EndOfBibitem
\bibitem[Ceriotti \latin{et~al.}(2010)Ceriotti, Bussi, and
  Parrinello]{ceri+10jctc}
Ceriotti,~M.; Bussi,~G.; Parrinello,~M. {Colored-Noise Thermostats {\`{a}} la
  Carte}. \emph{J. Chem. Theory Comput.} \textbf{2010}, \emph{6},
  1170--1180\relax
\mciteBstWouldAddEndPuncttrue
\mciteSetBstMidEndSepPunct{\mcitedefaultmidpunct}
{\mcitedefaultendpunct}{\mcitedefaultseppunct}\relax
\EndOfBibitem
\bibitem[Prabhakaran \latin{et~al.}(2012)Prabhakaran, Raman, Vogt, and
  Roth]{Prabhakaran2012}
Prabhakaran,~S.; Raman,~S.; Vogt,~J.~E.; Roth,~V. \emph{Lecture Notes in
  Computer Science}; Springer Berlin Heidelberg, 2012; pp 458--467\relax
\mciteBstWouldAddEndPuncttrue
\mciteSetBstMidEndSepPunct{\mcitedefaultmidpunct}
{\mcitedefaultendpunct}{\mcitedefaultseppunct}\relax
\EndOfBibitem
\bibitem[Ester \latin{et~al.}(1996)Ester, Kriegel, Sander, and
  Xiaowei]{ester1996density}
Ester,~M.; Kriegel,~H.; Sander,~J.; Xiaowei,~X. \emph{A density-based algorithm
  for discovering clusters in large spatial databases with noise}; 1996\relax
\mciteBstWouldAddEndPuncttrue
\mciteSetBstMidEndSepPunct{\mcitedefaultmidpunct}
{\mcitedefaultendpunct}{\mcitedefaultseppunct}\relax
\EndOfBibitem
\bibitem[Rodriguez and Laio(2014)Rodriguez, and Laio]{rodr-laio14science}
Rodriguez,~A.; Laio,~A. {Machine learning. Clustering by fast search and find
  of density peaks.} \emph{Science (New York, N.Y.)} \textbf{2014}, \emph{344},
  1492--1496\relax
\mciteBstWouldAddEndPuncttrue
\mciteSetBstMidEndSepPunct{\mcitedefaultmidpunct}
{\mcitedefaultendpunct}{\mcitedefaultseppunct}\relax
\EndOfBibitem
\bibitem[Scott(1992)]{scott+92kde}
Scott,~D. \emph{Multivariate density estimation : theory, practice, and
  visualization}; Wiley: New York, 1992\relax
\mciteBstWouldAddEndPuncttrue
\mciteSetBstMidEndSepPunct{\mcitedefaultmidpunct}
{\mcitedefaultendpunct}{\mcitedefaultseppunct}\relax
\EndOfBibitem
\bibitem[Bowman and Azzalini(1997)Bowman, and Azzalini]{bow+97app}
Bowman,~A.~W.; Azzalini,~A. \emph{Applied Smoothing Techniques for Data
  Analysis: The Kernel Approach with S-Plus Illustrations (Oxford Statistical
  Science Series)}; Oxford University Press, 1997\relax
\mciteBstWouldAddEndPuncttrue
\mciteSetBstMidEndSepPunct{\mcitedefaultmidpunct}
{\mcitedefaultendpunct}{\mcitedefaultseppunct}\relax
\EndOfBibitem
\bibitem[Chen \latin{et~al.}(2010)Chen, Wiesel, Eldar, and Hero]{chen+10ieee}
Chen,~Y.; Wiesel,~A.; Eldar,~Y.~C.; Hero,~A.~O. Shrinkage Algorithms for MMSE
  Covariance Estimation. \emph{IEEE Transactions on Signal Processing}
  \textbf{2010}, \emph{58}, 5016--5029\relax
\mciteBstWouldAddEndPuncttrue
\mciteSetBstMidEndSepPunct{\mcitedefaultmidpunct}
{\mcitedefaultendpunct}{\mcitedefaultseppunct}\relax
\EndOfBibitem
\bibitem[Roy and Vetterli(2007)Roy, and Vetterli]{roy+07espc}
Roy,~O.; Vetterli,~M. The effective rank: A measure of effective
  dimensionality. 15th European Signal Processing Conference. 2007; pp
  606--610\relax
\mciteBstWouldAddEndPuncttrue
\mciteSetBstMidEndSepPunct{\mcitedefaultmidpunct}
{\mcitedefaultendpunct}{\mcitedefaultseppunct}\relax
\EndOfBibitem
\bibitem[Vedaldi and Soatto(2008)Vedaldi, and Soatto]{vedaldi+08cv}
Vedaldi,~A.; Soatto,~S. In \emph{Computer Vision -- ECCV 2008: 10th European
  Conference on Computer Vision, Marseille, France, October 12-18, 2008,
  Proceedings, Part IV}; Forsyth,~D., Torr,~P., Zisserman,~A., Eds.; Springer
  Berlin Heidelberg: Berlin, Heidelberg, 2008; pp 705--718\relax
\mciteBstWouldAddEndPuncttrue
\mciteSetBstMidEndSepPunct{\mcitedefaultmidpunct}
{\mcitedefaultendpunct}{\mcitedefaultseppunct}\relax
\EndOfBibitem
\bibitem[Carreira-Perpinan(2000)]{carre00mode}
Carreira-Perpinan,~M.~A. Mode-finding for mixtures of Gaussian distributions.
  \emph{IEEE Transactions on Pattern Analysis and Machine Intelligence}
  \textbf{2000}, \emph{22}, 1318--1323\relax
\mciteBstWouldAddEndPuncttrue
\mciteSetBstMidEndSepPunct{\mcitedefaultmidpunct}
{\mcitedefaultendpunct}{\mcitedefaultseppunct}\relax
\EndOfBibitem
\bibitem[MacQueen \latin{et~al.}(1967)MacQueen, \latin{et~al.}
  others]{macq67kmeans}
others,, \latin{et~al.}  Some methods for classification and analysis of
  multivariate observations. Proceedings of the fifth Berkeley symposium on
  mathematical statistics and probability. 1967; pp 281--297\relax
\mciteBstWouldAddEndPuncttrue
\mciteSetBstMidEndSepPunct{\mcitedefaultmidpunct}
{\mcitedefaultendpunct}{\mcitedefaultseppunct}\relax
\EndOfBibitem
\bibitem[Ester \latin{et~al.}(1996)Ester, Kriegel, Sander, Xu, \latin{et~al.}
  others]{este96dbscan}
others,, \latin{et~al.}  A density-based algorithm for discovering clusters in
  large spatial databases with noise. Kdd. 1996; pp 226--231\relax
\mciteBstWouldAddEndPuncttrue
\mciteSetBstMidEndSepPunct{\mcitedefaultmidpunct}
{\mcitedefaultendpunct}{\mcitedefaultseppunct}\relax
\EndOfBibitem
\bibitem[Mardia and Jupp(2000)Mardia, and Jupp]{mard+00dirstat}
Mardia,~K.~V.; Jupp,~P.~E. Distributions on spheres. \emph{Directional
  Statistics} \textbf{2000}, 159--192\relax
\mciteBstWouldAddEndPuncttrue
\mciteSetBstMidEndSepPunct{\mcitedefaultmidpunct}
{\mcitedefaultendpunct}{\mcitedefaultseppunct}\relax
\EndOfBibitem
\bibitem[Mardia \latin{et~al.}(2008)Mardia, Hughes, Taylor, and
  Singh]{mard+08jcs}
Mardia,~K.~V.; Hughes,~G.; Taylor,~C.~C.; Singh,~H. A Multivariate Von Mises
  Distribution with Applications to Bioinformatics. \emph{The Canadian Journal
  of Statistics / La Revue Canadienne de Statistique} \textbf{2008}, \emph{36},
  99--109\relax
\mciteBstWouldAddEndPuncttrue
\mciteSetBstMidEndSepPunct{\mcitedefaultmidpunct}
{\mcitedefaultendpunct}{\mcitedefaultseppunct}\relax
\EndOfBibitem
\bibitem[Sra(2012)]{sra12cs}
Sra,~S. A short note on parameter approximation for von Mises-Fisher
  distributions: and a fast implementation of $I_s(x)$. \emph{Computational
  Statistics} \textbf{2012}, \emph{27}, 177--190\relax
\mciteBstWouldAddEndPuncttrue
\mciteSetBstMidEndSepPunct{\mcitedefaultmidpunct}
{\mcitedefaultendpunct}{\mcitedefaultseppunct}\relax
\EndOfBibitem
\bibitem[Efron(1979)]{efro79as}
Efron,~B. Bootstrap Methods: Another Look at the Jackknife. \emph{The Annals of
  Statistics} \textbf{1979}, \emph{7}, 1--26\relax
\mciteBstWouldAddEndPuncttrue
\mciteSetBstMidEndSepPunct{\mcitedefaultmidpunct}
{\mcitedefaultendpunct}{\mcitedefaultseppunct}\relax
\EndOfBibitem
\bibitem[Murtagh and Contreras(2012)Murtagh, and
  Contreras]{murt-cont12algorithms}
Murtagh,~F.; Contreras,~P. Algorithms for hierarchical clustering: an overview.
  \emph{Wiley Interdisciplinary Reviews: Data Mining and Knowledge Discovery}
  \textbf{2012}, \emph{2}, 86--97\relax
\mciteBstWouldAddEndPuncttrue
\mciteSetBstMidEndSepPunct{\mcitedefaultmidpunct}
{\mcitedefaultendpunct}{\mcitedefaultseppunct}\relax
\EndOfBibitem
\bibitem[De \latin{et~al.}(2017)De, Musil, Ingram, Baldauf, and
  Ceriotti]{de+17jci}
De,~S.; Musil,~F.; Ingram,~T.; Baldauf,~C.; Ceriotti,~M. Mapping and
  classifying molecules from a high-throughput structural database.
  \emph{Journal of Cheminformatics} \textbf{2017}, \emph{9}, 6\relax
\mciteBstWouldAddEndPuncttrue
\mciteSetBstMidEndSepPunct{\mcitedefaultmidpunct}
{\mcitedefaultendpunct}{\mcitedefaultseppunct}\relax
\EndOfBibitem
\bibitem[Ward~Jr(1963)]{ward63jasa}
Ward~Jr,~J.~H. Hierarchical grouping to optimize an objective function.
  \emph{Journal of the American statistical association} \textbf{1963},
  \emph{58}, 236--244\relax
\mciteBstWouldAddEndPuncttrue
\mciteSetBstMidEndSepPunct{\mcitedefaultmidpunct}
{\mcitedefaultendpunct}{\mcitedefaultseppunct}\relax
\EndOfBibitem
\bibitem[Wales and Doye(1997)Wales, and Doye]{walesdoye97jpca}
Wales,~D.~J.; Doye,~J. P.~K. Global Optimization by Basin-Hopping and the
  Lowest Energy Structures of Lennard-Jones Clusters Containing up to 110
  Atoms. \emph{The Journal of Physical Chemistry A} \textbf{1997}, \emph{101},
  5111--5116\relax
\mciteBstWouldAddEndPuncttrue
\mciteSetBstMidEndSepPunct{\mcitedefaultmidpunct}
{\mcitedefaultendpunct}{\mcitedefaultseppunct}\relax
\EndOfBibitem
\bibitem[Wales \latin{et~al.}(1998)Wales, Miller, and Walsh]{wales98nature}
Wales,~D.~J.; Miller,~M.~A.; Walsh,~T.~R. Archetypal energy landscapes.
  \emph{Nature} \textbf{1998}, \emph{394}, 758--760, Copyright - Copyright
  Macmillan Journals Ltd. Aug 20, 1998; Last updated - 2012-11-14; CODEN -
  NATUAS\relax
\mciteBstWouldAddEndPuncttrue
\mciteSetBstMidEndSepPunct{\mcitedefaultmidpunct}
{\mcitedefaultendpunct}{\mcitedefaultseppunct}\relax
\EndOfBibitem
\bibitem[Pártay \latin{et~al.}(2010)Pártay, Bartók, and Csányi]{lj38nested}
Pártay,~L.~B.; Bartók,~A.~P.; Csányi,~G. Efficient Sampling of Atomic
  Configurational Spaces. \emph{The Journal of Physical Chemistry B}
  \textbf{2010}, \emph{114}, 10502--10512, PMID: 20701382\relax
\mciteBstWouldAddEndPuncttrue
\mciteSetBstMidEndSepPunct{\mcitedefaultmidpunct}
{\mcitedefaultendpunct}{\mcitedefaultseppunct}\relax
\EndOfBibitem
\bibitem[Steinhardt \latin{et~al.}(1983)Steinhardt, Nelson, and
  Ronchetti]{stei+83prb}
Steinhardt,~P.~J.; Nelson,~D.~R.; Ronchetti,~M. {Bond-orientational order in
  liquids and glasses}. \emph{Phys. Rev. B} \textbf{1983}, \emph{28},
  784--805\relax
\mciteBstWouldAddEndPuncttrue
\mciteSetBstMidEndSepPunct{\mcitedefaultmidpunct}
{\mcitedefaultendpunct}{\mcitedefaultseppunct}\relax
\EndOfBibitem
\bibitem[Moroni \latin{et~al.}(2005)Moroni, ten Wolde, and Bolhuis]{moro+05prl}
Moroni,~D.; ten Wolde,~P.~R.; Bolhuis,~P.~G. Interplay between Structure and
  Size in a Critical Crystal Nucleus. \emph{Phys. Rev. Lett.} \textbf{2005},
  \emph{94}, 235703\relax
\mciteBstWouldAddEndPuncttrue
\mciteSetBstMidEndSepPunct{\mcitedefaultmidpunct}
{\mcitedefaultendpunct}{\mcitedefaultseppunct}\relax
\EndOfBibitem
\bibitem[Coasne \latin{et~al.}(2007)Coasne, Jain, Naamar, and
  Gubbins]{coas+07prb}
Coasne,~B.; Jain,~S.~K.; Naamar,~L.; Gubbins,~K.~E. Freezing of argon in
  ordered and disordered porous carbon. \emph{Phys. Rev. B} \textbf{2007},
  \emph{76}, 085416\relax
\mciteBstWouldAddEndPuncttrue
\mciteSetBstMidEndSepPunct{\mcitedefaultmidpunct}
{\mcitedefaultendpunct}{\mcitedefaultseppunct}\relax
\EndOfBibitem
\bibitem[Ogata(1992)]{ogat+92pra}
Ogata,~S. Monte Carlo simulation study of crystallization in rapidly
  supercooled one-component plasmas. \emph{Phys. Rev. A} \textbf{1992},
  \emph{45}, 1122--1134\relax
\mciteBstWouldAddEndPuncttrue
\mciteSetBstMidEndSepPunct{\mcitedefaultmidpunct}
{\mcitedefaultendpunct}{\mcitedefaultseppunct}\relax
\EndOfBibitem
\bibitem[Becker and Karplus(1997)Becker, and Karplus]{beck-karp97jcp}
Becker,~O.~M.; Karplus,~M. {The topology of multidimensional potential energy
  surfaces: Theory and application to peptide structure and kinetics}. \emph{J.
  Chem. Phys.} \textbf{1997}, \emph{106}, 1495\relax
\mciteBstWouldAddEndPuncttrue
\mciteSetBstMidEndSepPunct{\mcitedefaultmidpunct}
{\mcitedefaultendpunct}{\mcitedefaultseppunct}\relax
\EndOfBibitem
\bibitem[Mortenson \latin{et~al.}(2002)Mortenson, Evans, and Wales]{mort+02jcp}
Mortenson,~P.~N.; Evans,~D.~A.; Wales,~D.~J. {Energy landscapes of model
  polyalanines}. \emph{J. Chem. Phys.} \textbf{2002}, \emph{117}, 1363\relax
\mciteBstWouldAddEndPuncttrue
\mciteSetBstMidEndSepPunct{\mcitedefaultmidpunct}
{\mcitedefaultendpunct}{\mcitedefaultseppunct}\relax
\EndOfBibitem
\bibitem[Frishman and Argos(1995)Frishman, and Argos]{fris+95prot}
Frishman,~D.; Argos,~P. Knowledge-based protein secondary structure assignment.
  \emph{Proteins: Structure, Function, and Bioinformatics} \textbf{1995},
  \emph{23}, 566--579\relax
\mciteBstWouldAddEndPuncttrue
\mciteSetBstMidEndSepPunct{\mcitedefaultmidpunct}
{\mcitedefaultendpunct}{\mcitedefaultseppunct}\relax
\EndOfBibitem
\bibitem[Kabsch and Sander(1983)Kabsch, and Sander]{kabs+83bp}
Kabsch,~W.; Sander,~C. Dictionary of protein secondary structure: Pattern
  recognition of hydrogen-bonded and geometrical features. \emph{Biopolymers}
  \textbf{1983}, \emph{22}, 2577--2637\relax
\mciteBstWouldAddEndPuncttrue
\mciteSetBstMidEndSepPunct{\mcitedefaultmidpunct}
{\mcitedefaultendpunct}{\mcitedefaultseppunct}\relax
\EndOfBibitem
\bibitem[Hollingsworth \latin{et~al.}(2012)Hollingsworth, Lewis, Berkholz,
  Wong, and Karplus]{hollingsworth+12jmb}
Hollingsworth,~S.~A.; Lewis,~M.~C.; Berkholz,~D.~S.; Wong,~W.-K.;
  Karplus,~P.~A. $(\phi,\psi)_2$ Motifs: A Purely Conformation-Based
  Fine-Grained Enumeration of Protein Parts at the Two-Residue Level.
  \emph{Journal of Molecular Biology} \textbf{2012}, \emph{416}, 78 -- 93\relax
\mciteBstWouldAddEndPuncttrue
\mciteSetBstMidEndSepPunct{\mcitedefaultmidpunct}
{\mcitedefaultendpunct}{\mcitedefaultseppunct}\relax
\EndOfBibitem
\bibitem[Nagy and Oostenbrink(2014)Nagy, and Oostenbrink]{nagy+14jcim}
Nagy,~G.; Oostenbrink,~C. Dihedral-Based Segment Identification and
  Classification of Biopolymers I: Proteins. \emph{Journal of Chemical
  Information and Modeling} \textbf{2014}, \emph{54}, 266--277\relax
\mciteBstWouldAddEndPuncttrue
\mciteSetBstMidEndSepPunct{\mcitedefaultmidpunct}
{\mcitedefaultendpunct}{\mcitedefaultseppunct}\relax
\EndOfBibitem
\bibitem[Ardevol \latin{et~al.}(2015)Ardevol, Tribello, Ceriotti, and
  Parrinello]{arde+15jctc}
Ardevol,~A.; Tribello,~G.~A.; Ceriotti,~M.; Parrinello,~M. {Probing the
  Unfolded Configurations of a $\beta$-Hairpin Using Sketch-Map}. \emph{J.
  Chem. Theory Comput.} \textbf{2015}, \emph{11}, 1086--1093\relax
\mciteBstWouldAddEndPuncttrue
\mciteSetBstMidEndSepPunct{\mcitedefaultmidpunct}
{\mcitedefaultendpunct}{\mcitedefaultseppunct}\relax
\EndOfBibitem
\bibitem[Ramachandran \latin{et~al.}(1963)Ramachandran, Ramakrishnan, and
  Sasisekharan]{ramachandran1963stereochemistry}
Ramachandran,~G.~N.; Ramakrishnan,~C.; Sasisekharan,~V. Stereochemistry of
  polypeptide chain configurations. \emph{Journal of molecular biology}
  \textbf{1963}, \emph{7}, 95--99\relax
\mciteBstWouldAddEndPuncttrue
\mciteSetBstMidEndSepPunct{\mcitedefaultmidpunct}
{\mcitedefaultendpunct}{\mcitedefaultseppunct}\relax
\EndOfBibitem
\bibitem[Tompa \latin{et~al.}(2014)Tompa, Davey, Gibson, and
  Babu]{tompa+14molcell}
Tompa,~P.; Davey,~N.~E.; Gibson,~T.~J.; Babu,~M.~M. A million peptide motifs
  for the molecular biologist. \emph{Molecular cell} \textbf{2014}, \emph{55},
  161--169\relax
\mciteBstWouldAddEndPuncttrue
\mciteSetBstMidEndSepPunct{\mcitedefaultmidpunct}
{\mcitedefaultendpunct}{\mcitedefaultseppunct}\relax
\EndOfBibitem
\bibitem[Receveur-Br{\'e}chot and Durand(2012)Receveur-Br{\'e}chot, and
  Durand]{rece+12cpps}
Receveur-Br{\'e}chot,~V.; Durand,~D. How random are intrinsically disordered
  proteins? A small angle scattering perspective. \emph{Current Protein and
  Peptide Science} \textbf{2012}, \emph{13}, 55--75\relax
\mciteBstWouldAddEndPuncttrue
\mciteSetBstMidEndSepPunct{\mcitedefaultmidpunct}
{\mcitedefaultendpunct}{\mcitedefaultseppunct}\relax
\EndOfBibitem
\bibitem[Kruskal(1964)]{krusk64mds}
Kruskal,~J.~B. Multidimensional scaling by optimizing goodness of fit to a
  nonmetric hypothesis. \emph{Psychometrika} \textbf{1964}, \emph{29},
  1--27\relax
\mciteBstWouldAddEndPuncttrue
\mciteSetBstMidEndSepPunct{\mcitedefaultmidpunct}
{\mcitedefaultendpunct}{\mcitedefaultseppunct}\relax
\EndOfBibitem
\bibitem[Jolliffe(1986)]{joll86pca}
Jolliffe,~I.~T. \emph{Principal component analysis}; Springer, 1986; pp
  115--128\relax
\mciteBstWouldAddEndPuncttrue
\mciteSetBstMidEndSepPunct{\mcitedefaultmidpunct}
{\mcitedefaultendpunct}{\mcitedefaultseppunct}\relax
\EndOfBibitem
\bibitem[Abdi and Williams(2010)Abdi, and Williams]{abdi-lynn10pca}
Abdi,~H.; Williams,~L.~J. Principal component analysis. \emph{Wiley
  interdisciplinary reviews: computational statistics} \textbf{2010}, \emph{2},
  433--459\relax
\mciteBstWouldAddEndPuncttrue
\mciteSetBstMidEndSepPunct{\mcitedefaultmidpunct}
{\mcitedefaultendpunct}{\mcitedefaultseppunct}\relax
\EndOfBibitem
\bibitem[Coifman \latin{et~al.}(2005)Coifman, Lafon, Lee, Maggioni, Nadler,
  Warner, and Zucker]{coif05pnas}
Coifman,~R.~R.; Lafon,~S.; Lee,~A.~B.; Maggioni,~M.; Nadler,~B.; Warner,~F.;
  Zucker,~S.~W. Geometric diffusions as a tool for harmonic analysis and
  structure definition of data: Diffusion maps. \emph{Proceedings of the
  National Academy of Sciences of the United States of America} \textbf{2005},
  \emph{102}, 7426--7431\relax
\mciteBstWouldAddEndPuncttrue
\mciteSetBstMidEndSepPunct{\mcitedefaultmidpunct}
{\mcitedefaultendpunct}{\mcitedefaultseppunct}\relax
\EndOfBibitem
\bibitem[Maaten and Hinton(2008)Maaten, and Hinton]{maat08JMLR}
Maaten,~L. v.~d.; Hinton,~G. Visualizing data using t-SNE. \emph{Journal of
  Machine Learning Research} \textbf{2008}, \emph{9}, 2579--2605\relax
\mciteBstWouldAddEndPuncttrue
\mciteSetBstMidEndSepPunct{\mcitedefaultmidpunct}
{\mcitedefaultendpunct}{\mcitedefaultseppunct}\relax
\EndOfBibitem
\bibitem[Ceriotti \latin{et~al.}(2011)Ceriotti, Tribello, and
  Parrinello]{ceri+11pnas}
Ceriotti,~M.; Tribello,~G.~A.; Parrinello,~M. {From the Cover: Simplifying the
  representation of complex free-energy landscapes using sketch-map.}
  \emph{Proc. Natl. Acad. Sci. USA} \textbf{2011}, \emph{108}, 13023--8\relax
\mciteBstWouldAddEndPuncttrue
\mciteSetBstMidEndSepPunct{\mcitedefaultmidpunct}
{\mcitedefaultendpunct}{\mcitedefaultseppunct}\relax
\EndOfBibitem
\bibitem[Tribello \latin{et~al.}(2012)Tribello, Ceriotti, and
  Parrinello]{trib+12pnas}
Tribello,~G.~A.; Ceriotti,~M.; Parrinello,~M. {Using sketch-map coordinates to
  analyze and bias molecular dynamics simulations.} \emph{Proc. Natl. Acad.
  Sci. USA} \textbf{2012}, \emph{109}, 5196--201\relax
\mciteBstWouldAddEndPuncttrue
\mciteSetBstMidEndSepPunct{\mcitedefaultmidpunct}
{\mcitedefaultendpunct}{\mcitedefaultseppunct}\relax
\EndOfBibitem
\bibitem[Comitani \latin{et~al.}(2017)Comitani, Rossi, Ceriotti, Sanz, and
  Molteni]{comi+17jcp}
Comitani,~F.; Rossi,~K.; Ceriotti,~M.; Sanz,~M.~E.; Molteni,~C. {Mapping the
  conformational free energy of aspartic acid in the gas phase and in aqueous
  solution}. \emph{J. Chem. Phys.} \textbf{2017}, \emph{146}, 145102\relax
\mciteBstWouldAddEndPuncttrue
\mciteSetBstMidEndSepPunct{\mcitedefaultmidpunct}
{\mcitedefaultendpunct}{\mcitedefaultseppunct}\relax
\EndOfBibitem
\bibitem[Bonomi \latin{et~al.}(2008)Bonomi, Branduardi, Gervasio, and
  Parrinello]{bono+08jacs}
Bonomi,~M.; Branduardi,~D.; Gervasio,~F.~L.; Parrinello,~M. {The Unfolded
  Ensemble and Folding Mechanism of the C-Terminal GB1 $\beta$-Hairpin}.
  \emph{J. Chem. Am. Soc.} \textbf{2008}, \emph{130}, 13938--13944\relax
\mciteBstWouldAddEndPuncttrue
\mciteSetBstMidEndSepPunct{\mcitedefaultmidpunct}
{\mcitedefaultendpunct}{\mcitedefaultseppunct}\relax
\EndOfBibitem
\bibitem[Weber \latin{et~al.}(2004)Weber, Rungsarityotin, and
  Schliep]{webe+04pcca}
Weber,~M.; Rungsarityotin,~W.; Schliep,~A. \emph{Perron cluster analysis and
  its connection to graph partitioning for noisy data}; Konrad-Zuse-Zentrum
  f{\"u}r Informationstechnik Berlin, 2004\relax
\mciteBstWouldAddEndPuncttrue
\mciteSetBstMidEndSepPunct{\mcitedefaultmidpunct}
{\mcitedefaultendpunct}{\mcitedefaultseppunct}\relax
\EndOfBibitem
\bibitem[Deuflhard and Weber(2005)Deuflhard, and Weber]{deuf-webw05pcca}
Deuflhard,~P.; Weber,~M. Robust Perron cluster analysis in conformation
  dynamics. \emph{Linear algebra and its applications} \textbf{2005},
  \emph{398}, 161--184\relax
\mciteBstWouldAddEndPuncttrue
\mciteSetBstMidEndSepPunct{\mcitedefaultmidpunct}
{\mcitedefaultendpunct}{\mcitedefaultseppunct}\relax
\EndOfBibitem
\bibitem[R{\"o}blitz and Weber(2013)R{\"o}blitz, and Weber]{robl-webe13pcca+}
R{\"o}blitz,~S.; Weber,~M. Fuzzy spectral clustering by PCCA+: application to
  Markov state models and data classification. \emph{Advances in Data Analysis
  and Classification} \textbf{2013}, \emph{7}, 147--179\relax
\mciteBstWouldAddEndPuncttrue
\mciteSetBstMidEndSepPunct{\mcitedefaultmidpunct}
{\mcitedefaultendpunct}{\mcitedefaultseppunct}\relax
\EndOfBibitem
\bibitem[Chodera and No{\'e}(2014)Chodera, and No{\'e}]{chod-noe2014pcca}
Chodera,~J.~D.; No{\'e},~F. Markov state models of biomolecular conformational
  dynamics. \emph{Current opinion in structural biology} \textbf{2014},
  \emph{25}, 135--144\relax
\mciteBstWouldAddEndPuncttrue
\mciteSetBstMidEndSepPunct{\mcitedefaultmidpunct}
{\mcitedefaultendpunct}{\mcitedefaultseppunct}\relax
\EndOfBibitem
\bibitem[Valsson and Parrinello(2014)Valsson, and Parrinello]{vals-parr14prl}
Valsson,~O.; Parrinello,~M. {Variational Approach to Enhanced Sampling and Free
  Energy Calculations}. \emph{Phys. Rev. Lett.} \textbf{2014}, \emph{113},
  090601\relax
\mciteBstWouldAddEndPuncttrue
\mciteSetBstMidEndSepPunct{\mcitedefaultmidpunct}
{\mcitedefaultendpunct}{\mcitedefaultseppunct}\relax
\EndOfBibitem
\end{mcitethebibliography}
\end{document}